\documentclass[aip,
 amsmath,amssymb,
 reprint,%
]{revtex4-1}

\usepackage{graphicx}
\usepackage{dcolumn}
\usepackage{bm}

\usepackage[utf8]{inputenc}
\usepackage[T1]{fontenc}
\usepackage{mathptmx}
\usepackage{etoolbox}
\usepackage{xcolor}
\usepackage[nolist]{acronym}
\usepackage{tikz}
\usetikzlibrary{arrows.meta,positioning}
\usepackage{enumitem}
\usepackage[version=4]{mhchem}

\setlist[itemize]{
  leftmargin=*,
  itemsep=0pt,
  topsep=1pt,
  parsep=0pt,
  partopsep=0pt
}

\makeatletter
\def\@email#1#2{%
 \endgroup
 \patchcmd{\titleblock@produce}
  {\frontmatter@RRAPformat}
  {\frontmatter@RRAPformat{\produce@RRAP{*#1\href{mailto:#2}{#2}}}\frontmatter@RRAPformat}
  {}{}
}%

\makeatother
\begin{document}
\preprint{AIP/123-QED}

\title[An embedding method with constant potential boundary conditions]{An embedding method with constant potential boundary conditions}
\author{L. Hetzel}
\affiliation{ 
Department of Chemistry and Catalysis Research Center, TUM School of Natural Sciences, Technische Universität München, Lichtenbergstr. 4, Garching 85748, Germany 
}
\author{M. Head-Gordon}%

\affiliation{ 
Pitzer Center for Theoretical Chemistry, Department of Chemistry, University of California, Berkeley, California 94720, United States
}
\affiliation{ 
Chemical Sciences Division, Lawrence Berkeley National Laboratory, Berkeley, California 94720, United States}

\author{C. J. Stein}
\affiliation{ 
Department of Chemistry and Catalysis Research Center, TUM School of Natural Sciences, Technische Universität München, Lichtenbergstr. 4, Garching 85748, Germany 
}
\affiliation{%
Atomistic Modeling Center, Munich Data Science Institute, Technical University of Munich, Garching 85748, Germany
}%
 \email{christopher.stein@tum.de}

\date{\today}

\begin{abstract}
We present a self-consistent field framework for finite embedded quantum-chemical clusters with constant potential boundary conditions. The coupling is realized through an energy-independent self-energy commonly employed in quantum-transport calculations within the wide-band approximation. Starting from the corresponding non-equilibrium Green’s function formalism, we derive an analytic expression for the one-particle density matrix update that can be incorporated into conventional Hartree--Fock and density functional theory. The resulting non-Hermitian self-consistent field equations are solved using adapted Pulay-type mixing schemes. Applications to a quasi-periodic hydrogen ring demonstrate that a finite fragment coupled through an optimized self-energy accurately reproduces the polarization response of the extended system, while calculations on a lithium cluster capture metallic charge transfer and fractional occupations under open-boundary conditions. The proposed framework establishes practical grand-canonical boundary conditions for finite quantum-chemical clusters and lays the methodological foundation for quantum embedding methods for electrochemical systems.
\end{abstract}

\maketitle
\begin{acronym}
\acro{AIMD}{\textit{ab initio} molecular dynamics}
\acro{DFT}{Density Functional Theory}
\acro{RHF}{restricted Hartree--Fock}
\acro{HF}{Hartree--Fock}
\acro{SCF}{self-consistent field}
\acro{GC-DFT}{Grand-canonical density functional theory}
\acro{GC-SCF}{Grand-canonical self-consistent field}
\acro{NEGF}{non-equilibrium Green's function}
\acro{WBL}{wide-band limit}
\acro{AO}{atomic-orbital}
\acro{MO}{molecular orbital}
\acro{HOMO}{highest occupied molecular orbital}
\acro{LUMO}{lowest unoccupied molecular orbital}
\acro{MO}{molecular orbital}
\acro{SAD}{superposition of atomic densities}
\acro{RKS}{restricted Kohn--Sham}
\end{acronym}

\section{\label{sec:Intro} Introduction}

Electrocatalysis and electrochemistry play a central role in the conversion of electrical into chemical energy, enabling processes such as hydrogen fuel cells and solar-driven fuel production.\cite{li2020perspective,seh2017combining} From a theoretical perspective, these systems pose substantial challenges due to the intrinsic complexity of the electrochemical interface. Realistic models must account for an electrified electrode under bias, adsorbates, electrolyte, and solvent, whose collective interactions give rise to the electrochemical double layer. The resulting high-dimensional configuration space demands extensive sampling to obtain converged observables, leading to a fundamental trade-off between accuracy and computational efficiency.\cite{levell2024emerging}

A further fundamental challenge lies in the theoretical description of electrodes under an applied potential. Canonical electronic-structure methods employ a fixed number of electrons, whereas electrochemical conditions correspond to a grand-canonical ensemble, with the electrode acting as an electron reservoir at fixed potential. The appropriate thermodynamic potential (at temperature $0~$K) is therefore the grand potential,
\begin{equation}
\Omega = E - \mu N_e. \label{eq:grandpotential}
\end{equation}
where $E$ is the free energy, $\mu$ the chemical potential of the reservoir, and $N_e$ the number of electrons. In contrast to canonical electronic-structure methods, the electron number is no longer fixed but is determined by the reservoir chemical potential. \ac{GC-DFT} has been implemented in various forms\cite{ni2025gaussian,taylor2006first,jinnouchi2008electronic,lozovoi2001ab,otani2006first,letchworth2012joint,sundararaman2017grand,kastlunger2022using,bouzid2018atomic,beinlich2023controlled,hormann2020electrosorption,goodpaster2016identification} and successfully applied to study potential-dependent reaction energetics and kinetics, most prominently for CO\textsubscript{2} reduction (see, e.g., Refs.~\citenum{goodpaster2016identification,singh2017mechanistic,zhang2018importance,kastlunger2022using}). However, most existing grand-canonical approaches rely on periodic boundary conditions and require large supercells, making simulations of realistic electrochemical interfaces computationally demanding. Embedding approaches based on finite quantum-chemical clusters provide a natural route to reduce the computational cost of electrochemical simulations while potentially retaining an accurate description of the chemically active region.\cite{jones2020embedding,kolodzeiski2025efficient} Their application to electrochemical systems, however, requires grand-canonical boundary conditions that are applicable to finite cluster models. To address these limitations, we propose a grand-canonical embedding scheme for finite clusters. We first derive the working equations based on the equilibrium limit of quantum transport calculations of molecular junctions in a \acp{NEGF}\cite{arnold2007quantum} framework and then present an initial study of the corresponding numerical implementation for three model systems with both \ac{RHF} and \ac{DFT}.

\section{Theory}\label{sec:theory}
\subsection{Derivation of the working equations}
We start from a general expression for the density matrix based on a formalism derived for molecular junctions\cite{arnold2007quantum}, where we only consider a single lead, which is the electrode in our model:

\begin{eqnarray}\label{eq:P1}
   \mathbf{P} = \frac{1}{2\pi} \int_{-\infty}^\mu \text{d} E \mathbf{G}_R   \boldsymbol{\Gamma} \mathbf{G}_R^\dagger 
\end{eqnarray}
where $\mu$ is the chemical potential of the electrons and $\mathbf{G}_R$ are the retarded Green's functions in a given atomic orbital basis defined as

\begin{eqnarray}\label{eq:Gr}
   \mathbf{G}_R=(E\mathbf{S}-\mathbf{H}- \boldsymbol{\Sigma} )^{-1}.
\end{eqnarray}
with the Hamiltonian matrix $\mathbf{H}$, the energy $E$ and the overlap matrix $\mathbf{S}$. 
This constitutes an equilibrium limit of the molecular junction approach, where due to the absence of the second lead no chemical potential difference can be applied and electrons cannot flow through the junction, which then ultimately becomes the electrode (plus adsorbate) in our electrochemical model.
The energy broadening term is defined as
\begin{eqnarray}\label{eq:gamma}
    \boldsymbol{\Gamma}= i ( \boldsymbol{\Sigma}- \boldsymbol{\Sigma}^\dagger)\, ,
\end{eqnarray}
containing the self-energy $ \boldsymbol{\Sigma}$. We choose the self-energy to be diagonal and energy-independent, hence applying the \ac{WBL} (see Section~\ref{sec:theory}.B).

With the definition in Eq.~\ref{eq:Gr}, we can rewrite Eq. \ref{eq:P1} as (see Appendix \ref{app:eq:P_Gimg} for detailed derivation)
\begin{eqnarray}
  \mathbf{P} 
  &=& \frac{i}{2\pi}  \int_{-\infty}^\mu \text{d} E  \, \left(\mathbf{G_R}-\mathbf{G_R}^\dagger\right).\label{eq:P_Gimg}
\end{eqnarray} \\
To solve the integral in Eq. \ref{eq:P_Gimg} analytically, we transform to a basis in which $\mathbf{H}_\text{eff}$, the sum of the Hamiltonian and the self-energy, is diagonal

\begin{equation}\label{eq:B_eigenvalue}
    \underbrace{(\mathbf{H}+ \boldsymbol{\Sigma })}_{\mathbf{H}_\text{eff}}\mathbf{B}
    = \mathbf{B}\mathbf{Z}.
\end{equation}
Here, $\mathbf{B}$ is the transformation matrix containing the right eigenvectors and
$\mathbf{Z}$ is the diagonal matrix of complex eigenvalues, where we assume $\mathbf{H}_\text{eff}$ to be diagonalizable so that $\mathbf{B}$ is invertible.
We note that the self-energy is generally non-Hermitian and hence $\mathbf{H}_\text{eff}$ in Eq.~\ref{eq:B_eigenvalue} admits distinct left and right eigenvectors, which form a biorthogonal basis. Consequently, $\mathbf{B}$ is not generally unitary, such that $\mathbf{B}^{-1}\neq\mathbf{B}^{\dagger}$. Since the self-energy is chosen to be energy-independent, the eigenvalues and the transformation matrix are likewise energy-independent.

Solving the eigenvalue problem in Eq.~\ref{eq:B_eigenvalue} and assuming an orthonormal basis allows us to evaluate the integral in
Eq.~\ref{eq:P_Gimg} analytically. Accounting for the branch of the complex
logarithm yields the following expression for the density matrix,
\begin{eqnarray}
\mathbf{P}
&=&
\frac{i}{2\pi}
\Bigl[
\mathbf{B}\ln(\mathbf{Z}-\mu\mathbf{I})\mathbf{B}^{-1}
\notag\\
&&-
(\mathbf{B}^{-1})^\dagger
\ln(\mathbf{Z}^*-\mu\mathbf{I})
\mathbf{B}^\dagger
\Bigr].
\label{eq:density_final}
\end{eqnarray}
A detailed derivation is given in Appendix~\ref{app:deriv_final}, where
we also show that Eq.~\ref{eq:density_final} can equivalently be written
in a compact form directly in terms of $\mathbf{H}_{\mathrm{eff}}$ and
$\mathbf{H}_{\mathrm{eff}}^\dagger$. 
The obtained density matrix is independent of the form of the Hamiltonian matrix. Hence, it can be applied to standard electronic structure methods such as \ac{HF} or \ac{DFT}, which we both implemented and discuss in this article.
Similar models to the one introduced here have been widely adopted in quantum-transport calculations (see, e.g., Refs.~\citenum{arnold2007quantum,thoss2018perspective,datta2000nanoscale,cohen2020green}). For rigorous derivations of the underlying Keldysh formalism, we refer to the cited literature and standard textbooks.\cite{kadanoff_quantum_1962} 

\subsection{Model Parametrization}
We now discuss the parametrization of the self-energy and the role of the chemical potential. The self-energy $\boldsymbol{\Sigma}$ represents the thermodynamic reservoir and, in the context of electrochemical interfaces, is supposed to mimic a metallic electrode held at a constant bias potential $U_\text{bias}$. As noted in the previous section, we choose $\boldsymbol{\Sigma}$ to be diagonal in the atomic-orbital basis and energy independent, characterized simply by an orbital-dependent energy shift $\varepsilon_{pp}$ and broadening $\eta_{pp}$,
\begin{equation}
    \Sigma_{pp} = \varepsilon_{pp} - i \eta_{pp},
\end{equation}
where $\eta_{pp}>0$ for each orbital $p$. This choice corresponds to the \ac{WBL}, in which the density of states of the reservoir varies only weakly in the vicinity of the Fermi level and its influence can be captured by constant level shifts and broadenings.\cite{verzijl2013applicability,arnold2007quantum} A key advantage of this approximation is that it enables the analytic evaluation of Eq.~\ref{eq:P1} leading to the working equation~\ref{eq:density_final}. The \ac{WBL} has been shown to provide good agreement with fully self-consistent \ac{NEGF} calculations for bulk-metal electrodes, particularly when parts of the electrode are included explicitly, for example in the form of a finite cluster.\cite{verzijl2013applicability,covito2018transient} In practice, non-zero self-energy elements are restricted to orbitals directly coupled to the reservoir, such as valence orbitals on the boundary atoms of the metal cluster.\cite{arnold2007quantum}

The imaginary part $\eta_{pp}$ introduces finite lifetimes of the single-particle states and controls the rate of electron exchange with the reservoir, while the real part $\varepsilon_{pp}$ accounts for energy-level shifts induced by the coupling. The bath is further assumed to be time-independent, corresponding to a Markovian approximation with vanishing bath correlation time (see, e.g., Ref.~\citenum{RevModPhys.89.015001}). Electron exchange with the reservoir is thus treated as occurring on a much faster time scale than the electronic processes of interest. This assumption is well justified for macroscopic metallic electrodes with dense continua of states.\cite{arnold2007quantum}

In addition to the self-energy parameters, the chemical potential $\mu$ enters as a central control variable. It defines the upper integration limit in Eq.~\ref{eq:P1} and controls the electron number of the open system. Physically, $\mu$ corresponds to the Fermi level of the electronic reservoir and determines the charging state of the molecular subsystem or cluster model.\cite{kahn2016fermi} Setting the self energy to zero and choosing $\mu$ such that it lies between \ac{HOMO} and \ac{LUMO} recovers the canonical zero-temperature result.

\subsection{\ac{GC-SCF} algorithm}
\begin{figure}
\centering
\begin{tikzpicture}[
  box/.style={
    draw,
    rounded corners,
    align=center,
    inner sep=4pt,
    text width=7.6cm
  },
  arr/.style={-Latex, thick},
  node distance=4mm
]

\node[box] (init) {
\textbf{Inputs}\\
\begin{itemize}
    \item Chemical potential $\mu$ (can also be adapted during the cycle)
    \item Self-energy $\boldsymbol{\Sigma}$ in \ac{AO} basis
    \item Initial density $\mathbf{P}^{(0)}$ (e.g. from \ac{SAD} guess or canonical calculation)
\end{itemize}
};

\node[box, below=of init] (build) {
Build Fock matrix (HF/DFT)\\
$\mathbf{F}^{(n)} = \mathbf{H}_\mathrm{core} + \mathbf{V}[\mathbf{P}^{(n)}]$\\
$\mathbf{V}=\mathbf{V}_\mathrm{HF}$ (HF) or $\mathbf{V}_\mathrm{Hxc}$ (DFT)
};

\node[box, below=of build] (heff) {
Construct effective Hamiltonian\\
$\mathbf{H}_\mathrm{eff}^{(n)} = \mathbf{F}^{(n)} + \boldsymbol{\Sigma}$
};

\node[box, below=of heff] (orth) {
L\"owdin orthogonalization\\
$\mathbf{X}=\mathbf{S}^{-1/2}$,\quad
$\tilde{\mathbf{H}}_\mathrm{eff}=\mathbf{X}^\dagger \mathbf{H}_\mathrm{eff}\mathbf{X}$
};


\node[box, below=of orth] (dens) {
Grand-canonical density update\\
$\mathbf{P}_\mathrm{trial}^{(n+1)} = \dfrac{i}{\pi}\Bigl[\ln(\mathbf{H}_\mathrm{eff}^{(n)}-\mu\mathbf{I}) - \ln(\mathbf{H}_\mathrm{eff}^{(n)\dagger}-\mu\mathbf{I})\Bigr]$
};

\node[box, below=of dens] (mix) {
Update density Matrix using Pulay mixing
};

\node[box, below=of mix] (outputs) {
{\footnotesize
\textbf{Compute outputs}\\[3pt]
For HF: $E_\text{elec} = 0.5\cdot Tr[\mathbf{P}\cdot (\mathbf{H}_\mathrm{core} + \mathbf{F})]$\\[4pt]
For DFT: $\begin{aligned}[t]
    E_\text{elec} = &\ 0.5\cdot Tr[\mathbf{P}\cdot(\mathbf{H}_\mathrm{core}+\mathbf{F})]\\
        &+ E_\mathrm{xc}[\mathbf{P}] - 0.5\cdot Tr[\mathbf{P}\cdot\mathbf{V}_\mathrm{xc}]
    \end{aligned}$\\[4pt]
$N_e = Tr[\mathbf{P}\cdot \mathbf{S}]$
}
};

\node[box, below=of outputs] (conv) {
Convergence check};

\draw[arr] (init) -- (build);
\draw[arr] (build) -- (heff);
\draw[arr] (heff) -- (orth);
\draw[arr] (orth) -- (dens);
\draw[arr] (dens) -- (mix);
\draw[arr] (mix) -- (outputs);
\draw[arr] (outputs) -- (conv);

\draw[arr] (conv.east) -- ++(6mm,0) |- (build.east);

\end{tikzpicture}
\caption{Schematic overview of the restricted \ac{GC-SCF} workflow used in this work, similar to the workflow in Ref. \citenum{arnold2007quantum}.}
\label{fig:gcsf_flow}
\end{figure}
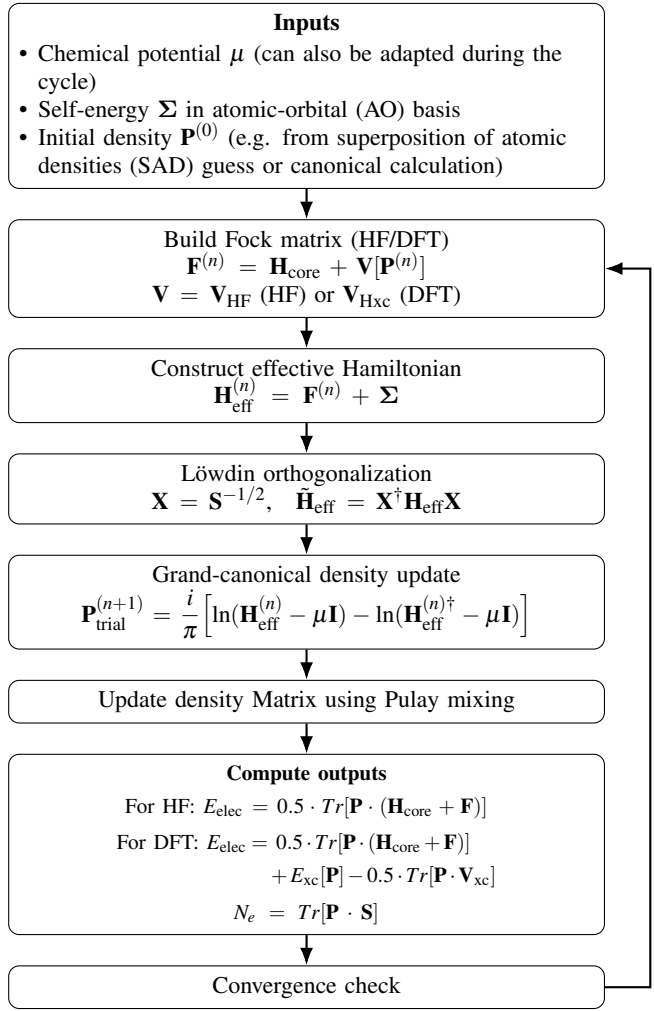

The described framework can, in principle, be implemented for any mean-field method. 
We present here both a restricted \ac{HF} and a \ac{DFT} implementation. 
In this section, we describe the iterative procedure used to obtain a self-consistent solution. 
The \ac{GC-SCF} scheme seeks a fixed point of the nonlinear map defined by Eq.~\ref{eq:density_final},
i.e., the density matrix is updated until convergence is reached. 
The corresponding workflow is shown in Fig.~\ref{fig:gcsf_flow}.

The calculation is initialized by specifying the chemical potential $\mu$ and the self-energy $\boldsymbol{\Sigma}$, together with an initial density matrix $\mathbf{P}^{(0)}$. From this density, the Fock matrix is constructed as $\mathbf{F}^{(n)} = \mathbf{H}_\mathrm{core} + \mathbf{V}[\mathbf{P}^{(n)}]$, where $\mathbf{H}_\mathrm{core}$ contains the usual one-electron kinetic energy and electron–nuclear attraction terms. The effective Hamiltonian is then formed by adding the self-energy, whose non-zero diagonal elements are restricted to the coupled atomic orbitals in the \ac{AO} basis.
Subsequently, the generalized eigenvalue problem is transformed to an
orthogonal basis via symmetric L\"owdin orthogonalization for convenience.
The density matrix is then obtained directly from the restricted form of Eq.~\ref{eq:P_log_Heff}, without requiring an explicit
diagonalization of $\mathbf{H}_{\mathrm{eff}}$. The presence of the bath, coupled at a constant chemical potential through a self-energy with finite broadening, leads to a non-idempotent density matrix and hence an ensemble interpretation of the associated quantum state at fixed chemical potential and choice of self-energy parameters.
The resulting fixed-point problem is therefore solved iteratively using
Pulay-type\cite{pulay1980convergence} mixing schemes until convergence is
achieved.

\section{Computational methods}

If not mentioned otherwise, all electronic-structure calculations were performed using the PySCF package (version~2.13.0)~\cite{sun2015libcint,sun2018pyscf,sun2020_pyscf}, which was used to evaluate one- and two-electron integrals and to carry out canonical \ac{HF} and \ac{RKS} \ac{DFT} calculations. The \ac{GC-SCF} framework was implemented as a custom Python package interfaced with PySCF. The implementation supports several Pulay-type mixing schemes~\cite{pulay1980convergence,pulay1982improved}, adapted to the grand-canonical setting. An overview of the implemented mixing algorithms, together with the corresponding mixed quantities and residuals or target functionals, is given in Table~\ref{tab:diis}.

\begin{table}
\caption{\label{tab:diis}
Overview of mixing schemes implemented in the grand-canonical SCF framework, indicating the mixed quantity and the corresponding residual or target functional.}
\begin{ruledtabular}
\begin{tabular}{lccc}
Name & Mixed quantity & Residual / target & Ref.\\
\hline
CDIIS
& Fock matrix $\mathbf{F}$
& $\mathbf{F}\mathbf{P}\mathbf{S}-\mathbf{S}\mathbf{P}\mathbf{F}$
& \citenum{pulay1980convergence,pulay1982improved}\\

EDIIS(E)
& Fock matrix $\mathbf{F}$
& $E[\mathbf{P}]$
& \citenum{kudin2002black}\\

EDIIS($\Omega$)
& Fock matrix $\mathbf{F}$
& $\Omega[\mathbf{P}]$
& this work\\

Anderson
& Density matrix $\mathbf{P}$
& $  \mathbf{P}_\mathrm{out} - \mathbf{P}_\mathrm{in}$
& \citenum{anderson1965iterative}\\
\end{tabular}
\end{ruledtabular}
\end{table}

In addition to the standard schemes, we implemented an adapted EDIIS variant using the grand potential (see Eq.~\ref{eq:grandpotential}) as the target functional instead of the total energy. Moreover, a hybrid scheme combining EDIIS($\Omega$) and Anderson mixing, denoted EDIIS($\Omega$)+A, was implemented. In this approach, EDIIS($\Omega$) is employed during the initial iterations to provide robust convergence, and the algorithm switches to Anderson mixing only after at least 12 SCF iterations and once the density-matrix change satisfies $\|\Delta \mathbf{P}\| < 10^{-2}$. To improve numerical stability, accelerated Anderson steps were rejected whenever the grand potential of the accelerated density exceeded that of the corresponding unaccelerated fixed-point step. In such cases, the unaccelerated trial density was accepted instead. More details on the implementation are provided in the SI (see section S1).

\paragraph{Single water molecule}

For initial convergence tests with a single water molecule, we employed the minimal atomic orbital (MINAO) basis set as implemented in PySCF and the STO-3G\cite{hehre1969self} basis set. Calculations were performed both with \ac{RHF} and Kohn–Sham \ac{DFT}, using the PBE\cite{perdew1996generalized} exchange–correlation functional for the latter. The self-energy was applied to the oxygen $2s$ and $2p$, as well as the hydrogen $1s$ valence orbitals; all remaining $\boldsymbol{\Sigma}$ matrix elements were set to zero. For each basis-set and method combination, all Pulay mixing schemes discussed above (see Table~\ref{tab:diis}), as well as EDIIS($\Omega$)+A, were tested, starting from both a \ac{SAD} initial guess and from a converged canonical ($\boldsymbol{\Sigma}=0$) calculation as the initial density guess. The DIIS subspace size was set to 12 for all calculations, and a maximum of 200 SCF iterations was allowed. All calculations were performed for the neutral, closed-shell water molecule (see SI section S2.1 for the coordinates). \ac{SCF} convergence was assessed using the density-matrix change and, for grand-canonical calculations, additionally the change in the grand potential, requiring the Frobenius norm $\|\Delta \mathbf{P}\| < 10^{-6}$ and $|\Delta \Omega| < 10^{-8}$.

\paragraph{Hydrogen ring}
The reference system comprised 40 hydrogen atoms with a bond distance of 0.74~\AA~ arranged on a ring. The coordinates are provided in the SI (see section S3.1). The system was treated at the \ac{RKS}--\ac{DFT} level using the hybrid exchange--correlation functional PBE0\cite{adamo1999toward} in combination with the MINAO basis set. The canonical single-point calculations were calculated using pySCF's default settings. 

The fragment treated with our \ac{GC-SCF} implementation consisted of six to twenty hydrogen atoms (in increments of two), arranged in the same ring-like structure. These fragments were treated using the same level of theory as the reference, with Anderson mixing to converge the \ac{SCF}. We coupled only the 1\textit{s} orbitals of the outer hydrogen atoms to the reservoir, i.e., all other entries of the self-energy are set to zero. The required parameters were optimized to reproduce the reference. The optimization was conducted with the L-BFGS\cite{liu1989limited} algorithm, terminated when either the relative
reduction in the objective function fell below $\mathrm{ftol} = 10^{-8}$ or the projected gradient norm fell below
$\mathrm{gtol} = 10^{-5}$. At each optimization step, the chemical potential was self-consistently adjusted to enforce charge neutrality of the fragment in the absence of the external point charge. More information on the optimization strategy is provided in the SI (see section S3.2).

\paragraph{Lithium cluster}
A \ce{Li40} cluster was constructed from a Li BCC(110) surface using the experimental lattice constant of 3.51~\AA.\cite{beg1976temperature} The coordinates are provided in the SI (see Section~S4.1). Electronic-structure calculations were performed using the PBE0 exchange--correlation functional\cite{adamo1999toward} in combination with the STO-3G basis set.\cite{hehre1969self} Grand-canonical calculations were carried out using the \ac{GC-SCF} framework with Anderson mixing and $\varepsilon = 0$. The self-energy was applied exclusively to the 2\textit{s} valence orbitals of selected lithium atoms, where all remaining $\boldsymbol{\Sigma}$ matrix elements of the self-energy were set to zero. Three coupling topologies were considered: a center scheme coupling interior bulk-like atoms, a surface-center scheme, and an edges scheme coupling the peripheral atoms of the cluster (see SI section S4.1). For the broadening scan, $\eta$ was varied across several orders of magnitude from $0.01~\text{mHa}$ to $500~\text{mHa}$, with the chemical potential self-consistently adjusted to enforce charge neutrality. For the charge scan, a positive point charge $q = +0.4\,e$ was moved along the surface normal above the center top atom with a moderate broadening of $\eta = 2.0~\text{mHa}$ and the chemical potential fixed to the value that ensures charge neutrality in the unperturbed cluster. A canonical reference calculation was performed at the same level of theory.

\section{Results and Discussion}
\subsection{Validation on a water molecule} \label{sec:water}

\begin{figure*}
\includegraphics[width=0.9\textwidth]{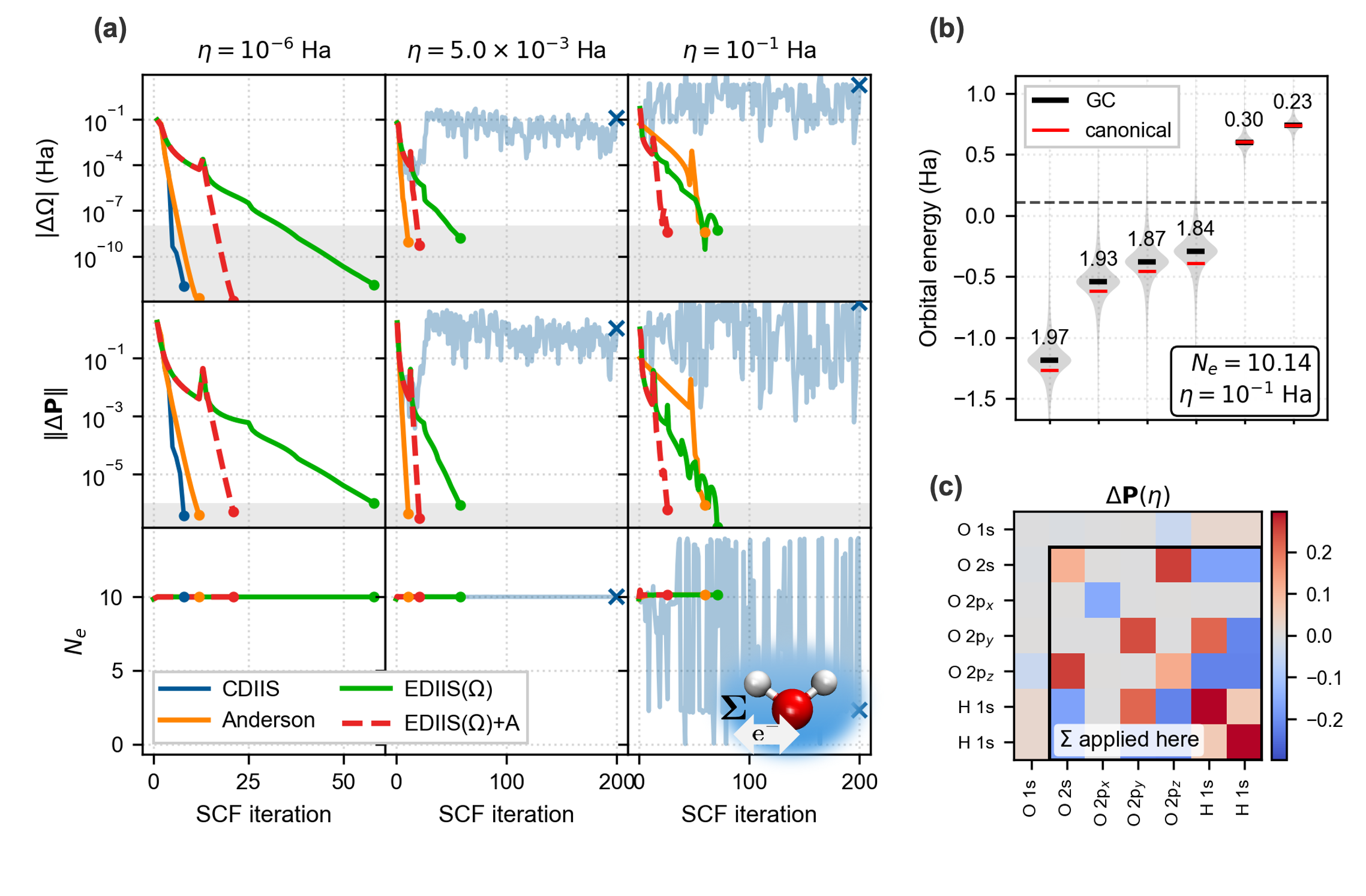}
\caption{
\textbf{(a)} Convergence behavior of the grand-canonical SCF procedure for three broadening strengths $\eta$ (columns), comparing CDIIS, EDIIS($\Omega$), Anderson, and a hybrid EDIIS($\Omega$)+Anderson scheme, starting from an \ac{SAD} guess. Shown are the iteration-wise changes in the grand potential, $|\Delta \Omega|$, and the Frobenius norm of the density-matrix update, $\|\Delta \mathbf{P}\|$. Grey shaded regions indicate the respective convergence thresholds. The light-blue curve in the right panels indicates the unconverged CDIIS calculation.
\textbf{(b)} Molecular orbital (MO) energies from the converged grand-canonical calculation at $\eta=10^{-1}\,\mathrm{Ha}$. Horizontal lines denote MO eigenvalues. Shaded regions illustrate a schematic spectral broadening centered at each \ac{MO}, with widths proportional to the leakage parameter $\eta$ weighted by the \ac{AO} character of the corresponding MO. Grand-canonical orbital occupations are indicated numerically. The chemical potential $\mu$ is shown as a dashed horizontal line.
\textbf{(c)} Change in the density matrix $\Delta\mathbf{P} = \mathbf{P}(\eta) - \mathbf{P}(\eta=0)$ shown in the \ac{AO} basis for $\eta=10^{-1}\,\mathrm{Ha}$.
\label{fig:water}
}
\end{figure*}

As a first validation of the \ac{GC-SCF} framework,
we consider a single water molecule as a simple, finite reference system. First, we examine the convergence behavior of the grand-canonical SCF procedure for several values of $\eta$ using different Pulay-type mixing schemes. Unless stated otherwise, the results shown in the main text are obtained at the
\ac{RHF} level using the STO-3G basis set.
Figure \ref{fig:water} illustrates the iterative convergence of the grand potential $\Omega$, the electron number $N_e$, and the density-matrix update for representative broadening strengths $\eta$ and different mixing schemes. To assess whether different mixing schemes converge to the same grand-canonical fixed point, Table \ref{tab:water_fixedpoint_consistency} reports the final converged values of $\Omega$, $N_e$, and the Frobenius norm of the Fock–density commutator residual for the strongest broadening $\eta = 10^{-1}$ Ha, starting from two distinct initial density guesses.

\begin{table*}
\caption{
Final grand-canonical SCF results for a single water molecule
at the RHF/STO-3G level for a strong imaginary broadening
$\eta~=~1\times10^{-1}\,\mathrm{Ha}$.
Reported are the converged grand potential $\Omega$,
the electron number $N_e$, the Frobenius norm of the
Fock--density commutator residual
$\|\mathbf{R}\| = \|[\mathbf{F},\mathbf{P}\mathbf{S}]\|$ and the number of iterations until convergence $N_\mathrm{iter}$.
Results are shown for two distinct initial density guesses:
a \ac{SAD} initial guess and a density
obtained from a converged canonical ($\boldsymbol{\Sigma}=0$) RHF calculation.
}
\label{tab:water_fixedpoint_consistency}
\centering
\begin{ruledtabular}
\begin{tabular}{l|cccc|cccc}
 & \multicolumn{4}{c}{\ac{SAD} guess}
 & \multicolumn{4}{c}{$\boldsymbol{\Sigma}=0$ start} \\
 SCF scheme
 & $\Omega$ (Ha) & $N_e$ & $\|\mathbf{R}\|$ & $N_\mathrm{iter}$
 & $\Omega$ (Ha) & $N_e$ & $\|\mathbf{R}\|$ & $N_\mathrm{iter}$ \\
\hline
CDIIS & -53.065574 & 2.3050 & -- & -- & -75.523593 & 10.1913 & $1.24\times10^{-1}$ & 10 \\
EDIIS($\Omega$) & -75.496810 & 10.1389 & $1.19\times10^{-1}$ & 72 & -75.496810 & 10.1389 & $1.19\times10^{-1}$ & 65 \\
Anderson & -75.496810 & 10.1389 & $1.19\times10^{-1}$ & 61 & -75.496810 & 10.1389 & $1.19\times10^{-1}$ & 123 \\
EDIIS($\Omega$)+A & -75.496810 & 10.1389 & $1.19\times10^{-1}$ & 27 & -75.496810 & 10.1389 & $1.19\times10^{-1}$ & 43 \\
\end{tabular}

\end{ruledtabular}
\end{table*}
With the exception of conventional CDIIS, all mixing schemes converge to identical values of $\Omega$ and $N_e$ within numerical precision, independent of the initial density guess. In particular, Anderson mixing, EDIIS($\Omega$), and the hybrid scheme EDIIS($\Omega$)+A consistently reach the same grand-canonical fixed point, confirming that the solution is well-defined and algorithm-independent even in the strongly broadened regime. Since the chemical potential is fixed throughout the calculation and the number of electrons barely changes, EDIIS(E) and EDIIS($\Omega$) are indistinguishable. The results for EDIIS(E) are therefore reported only in the SI (see Table S1--S4). The failure of CDIIS for large $\eta$ is expected, as the effective Hamiltonian becomes increasingly non-Hermitian, rendering the conventional commutator residual ill-suited as a convergence measure. As shown in Fig.~\ref{fig:water}a, the hybrid scheme yields the fastest convergence for this system. Additional tests using other basis sets and \ac{DFT} are provided in the SI in Section~S2. For \ac{DFT} calculations, we obtain similar trends in convergence behavior. We therefore identify EDIIS($\Omega$) and Anderson as robust fallback strategies, while the hybrid scheme may be employed as an efficient default choice.

In addition to the convergence behavior, we analyze how the imaginary self-energy modifies the electronic structure at the orbital and density-matrix levels. Figure~\ref{fig:water}b shows the \ac{MO} energies obtained from the converged canonical and grand-canonical calculations for a strong broadening strength of $\eta = 10^{-1}\,\mathrm{Ha}$. The effect of the imaginary self-energy is illustrated schematically by broadening profiles centered at the \ac{MO} eigenvalues. As a result, the introduction of the imaginary self-energy leads to fractional orbital occupations differing substantially from 0 or 2 for orbitals near the chemical potential, which was chosen at the midpoint of the canonical HOMO--LUMO gap. 
Additionally, the energy levels of the grand-canonical calculation are slightly shifted compared to the canonical reference. 
Although the \acp{MO} are not identical for the canonical and grand-canonical case, the comparison illustrates the overall redistribution of electronic occupation induced by the reservoir coupling. 
The effect of the imaginary self-energy on the electronic structure is further illustrated in Fig.~\ref{fig:water}c, which shows the change in the one-particle reduced density matrix relative to the canonical calculation evaluated in the atomic-orbital basis. 
Since the self-energy is applied to all valence orbitals of the water molecule, the largest modifications occur in matrix elements associated with the coupled valence orbitals, while the deeply bound oxygen \textit{1s} orbital exhibits only slight changes, consistent with zero coupling. 

We note that the isolated water molecule was intentionally chosen here as a simple, closed-shell reference system. The present results primarily validate the numerical stability and fixed-point consistency of the grand-canonical SCF framework. Applications to more physically relevant systems are discussed in the following sections.

\begin{figure*}[t!]
\includegraphics[width=0.9\textwidth]{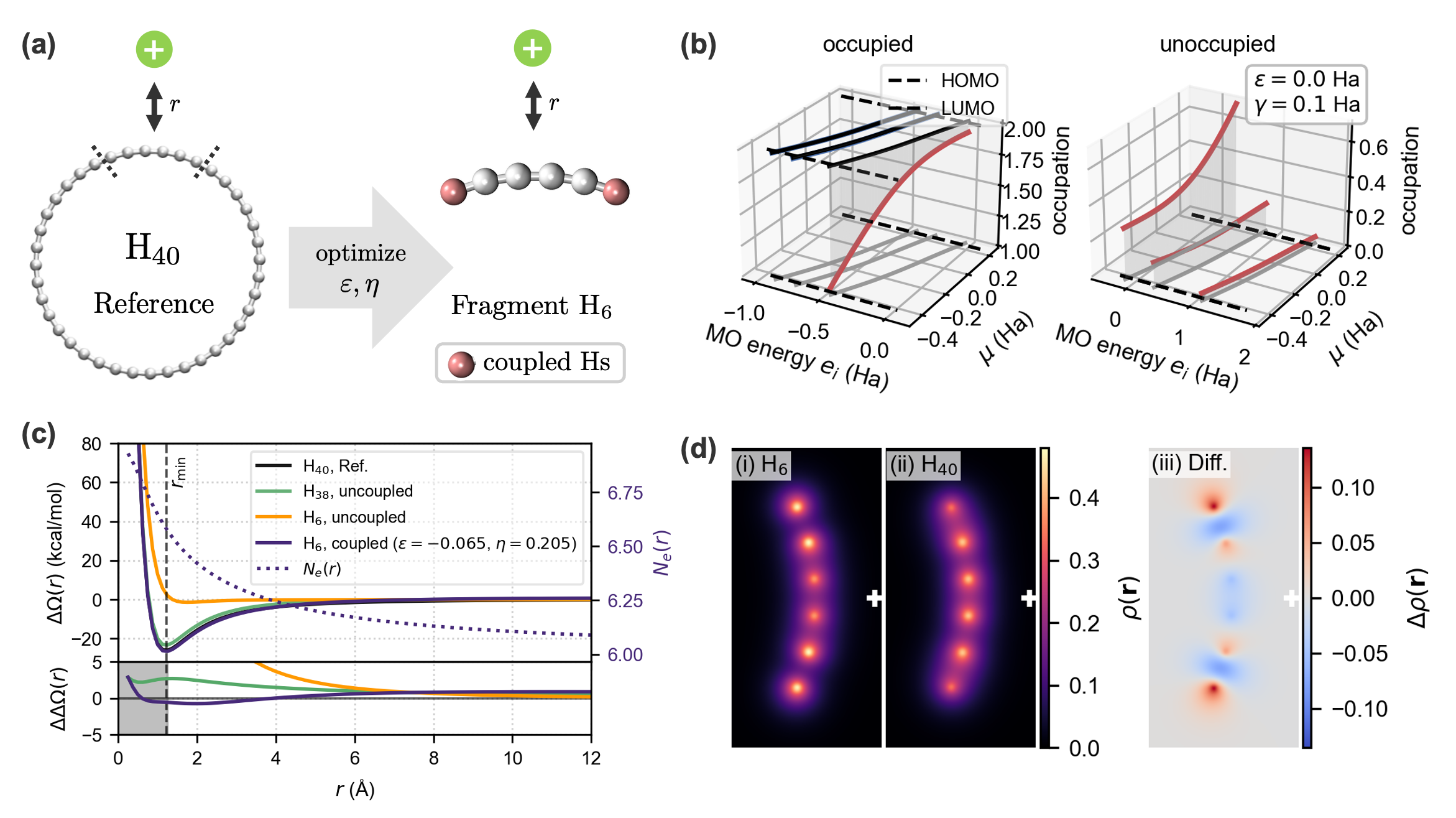}
\caption{
\textbf{(a)} Schematic illustration of the 1D quasi-periodic system. The \ce{H40} ring serves as the reference for the \ce{H6} fragment, where the outer hydrogen atoms highlighted in red are coupled to the reservoir. The reaction coordinate $r$ is defined by the distance of the point charge and the closest two hydrogen atoms.
\textbf{(b)} \ac{MO} energies $e_i$ and grand-canonical occupations as a function of the chemical potential $\mu$ for the fragment coupled to a self-energy with parameters $(\varepsilon = 0,\, \eta = 0.1~\mathrm{Ha})$. The left panel shows the three occupied and the right panel the three unoccupied orbitals. Red curves indicate orbitals whose weight in the coupled subspace exceeds $15\%$, defined as
$w_i = \sum_{\nu \in \Sigma} |C_{\nu i}|^2,$ where $\Sigma$ denotes the atomic orbitals entering the self-energy and $C_{\nu i}$ are the molecular orbital coefficients.
\textbf{(c)} Potential curve of the point charge approaching the hydrogen system for the uncoupled fragment (blue), the reference system, and the coupled systems with the optimized parameters shown in the upper panel, where $\Delta \Omega=\Omega(r)-\Omega_0$, where $\Omega_0$ is the grand potential of the fragment in the absence of a positive point charge. The lower panel shows the residual relative to the \ce{H40} reference system in kcal/mol. The obtained chemical potential fulfilling the neutrality constraint is $-0.2884$ Ha.
\textbf{(d)}
Two-dimensional $xy$-plane ($z=0$) electron-density slices at the adsorption minimum ($r=1.22~\text{\AA}$) for an external point charge $q=+1$ placed along the ring axis (white cross).
Panel (i) shows the grand-canonical fragment density $\rho_{\mathrm{frag}}(\mathbf{r})$ for \ce{H6} coupled to the self-energy using $(\varepsilon,\eta)$ and the chemical potential $\mu$ obtained from the coupled calculation.
Panel (ii) shows the density $\rho_{\mathrm{ref.}}(\mathbf{r})$ obtained by projecting the full H$_{40}$ density matrix onto the atomic-orbital subspace of the six hydrogen atoms closest to the charge.
Panel (iii) shows the real density difference $\Delta\rho(\mathbf{r})=\rho_{\mathrm{frag}}(\mathbf{r})-\rho_{\mathrm{ref.}}(\mathbf{r})$. All underlying calculations were conducted using PBE0 and a minimal basis (MINAO).
\label{fig:hring}
}
\end{figure*}

\subsection{1D quasi-periodic system}\vspace{-0.5cm}
In this section, we turn to a one-dimensional quasi-periodic system that exhibits near-metallic behavior. Specifically, we consider a \ce{H40} hydrogen chain arranged in a ring geometry that is approached by a positive point charge, thereby inducing polarization effects. 
The aim is to employ a minimal \ce{H6} fragment which is adequately coupled within the \ac{GC-SCF} framework that reproduces the polarization-induced adsorption energy of the point charge on the full system \ce{H40} using a minimal set of self-energy parameters. 
The model system is shown in Fig.~\ref{fig:hring}a. In Fig.~\ref{fig:hring}b, we analyze the $\mu$-dependence of the fragment within the \ac{GC-SCF} framework, where the \ac{MO} energies $e_i$ are shown together with their grand-canonical occupations. For $\eta = 0.1~\mathrm{Ha}$ and $\varepsilon = 0$, the largest occupation changes occur in the vicinity of the HOMO and LUMO. As the chemical potential is varied across the frontier region, these orbitals exhibit the expected continuous occupation transfer, while deeper occupied and higher virtual states remain essentially unaffected. 
Importantly, the HOMO–LUMO gap remains nearly constant throughout the scan despite substantial changes in the orbital energies, indicating that the coupling mainly induces a redistribution of electron density rather than a qualitative change in the orbital spectrum.

In the next step, we seek to reproduce the response of the reference system to a constant charge with an optimized set of self-energy parameters. 
The parameters were determined by minimizing a loss function that enforces agreement between the potential curve of a \ce{H6} fragment and that of the full ring. At each optimization step, the chemical potential is self-consistently adjusted to enforce charge neutrality of the fragment in the absence of the external point charge, thereby ensuring that the optimized response is not driven by a spurious net charge transfer between the fragment and the reservoir. Further technical details of the optimization scheme are provided in the SI (see section S3.2). To assess whether the optimization landscape contains competing local minima, we performed a grid scan over the coupling parameters, shown in the SI (see Fig.~S9). Around the primary optimum, the loss function exhibits a broad, shallow minimum along $\eta$, indicating that the fit constrains the broadening only weakly within this range. A second local minimum occurs at $\varepsilon=-0.060$~Ha and $\eta=0.329$~Ha, corresponding to a broadening approximately 60\% larger than at the primary minimum, while $\varepsilon$ changes only marginally. Despite this substantial difference in $\eta$, the resulting potential energy curve is nearly indistinguishable from that obtained at the primary minimum (see Fig.~S10a). The corresponding electron density, however, shows larger deviations from the reference ring (see Fig.~S10b). We therefore retain the primary minimum for the subsequent analysis.

The resulting potential energy curve using the optimized parameters is shown in Fig.~\ref{fig:hring}c. For comparison, we also include the uncoupled \ce{H6} fragment and a ring comprising 38 hydrogen atoms, which is similar in size to the reference system but breaks the periodicity.
The uncoupled fragment lacks an attractive interaction and remains purely repulsive. In contrast, the \ce{H6}
fragment with optimized coupling parameters reproduces the reference system's well depth closely, with a
deviation of approximately 0.57~kcal/mol or 2.2\% at the minimum $r_\mathrm{min}$. For the \ce{H38} ring,
however, a deviation of approximately 2.7~kcal/mol or 10.3\% is obtained at the minimum $r_\mathrm{min}$,
indicating that even a small perturbation of the periodicity significantly alters the polarization response
of the system.

Although the coupled fragment reproduces the polarization response with high accuracy, residual deviations of up to $\sim 0.9~\mathrm{kcal/mol}$ remain in the intermediate and tail region between the minimum and the asymptotic approach to zero. The projected electron densities and density differences shown in Fig.~\ref{fig:hring}d reveal that the local response in the vicinity of the point charge is in good agreement with the reference system, whereas noticeable deviations occur at the fragment boundaries. In particular, the induced density is more localized within the fragment, and the extended polarization present in the full ring is partially truncated.
This suggests that the deviations in the intermediate region originate from finite-size effects, as the polarization cloud cannot fully extend beyond the fragment boundaries.
In contrast, the \ce{H38} ring exhibits deviations of the opposite sign near $r_\mathrm{min}$, indicating that this intermediate region is sensitive not only to finite-size truncation but can also not be reproduced when the periodicity is broken. 

\begin{figure}
\includegraphics{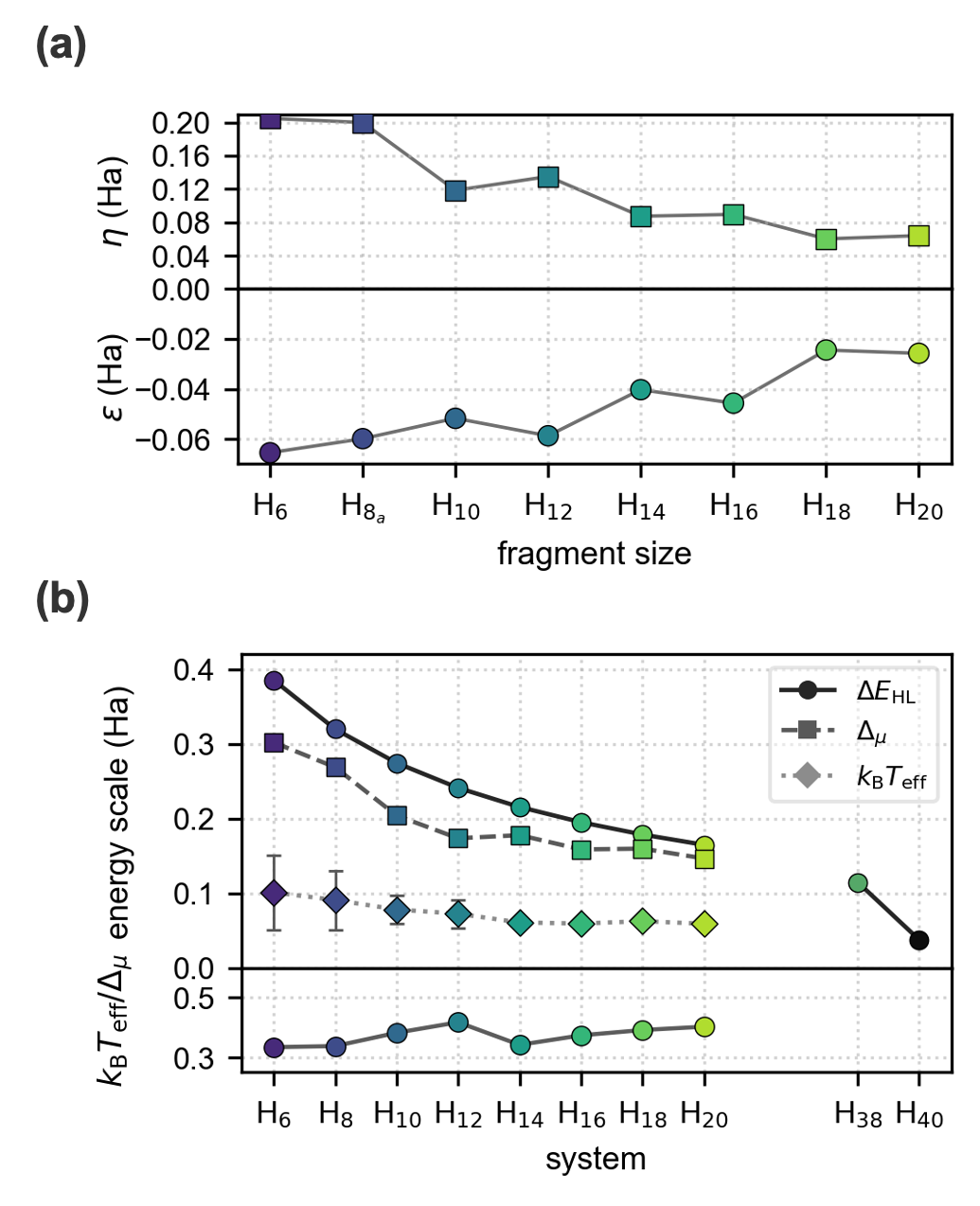}
\caption{
\textbf{(a)}  Optimized self-energy parameters as a function of fragment size. \textbf{(b)} Electronic-structure properties of the optimized hydrogen-ring fragments and reference systems. Upper panel: canonical HOMO--LUMO gaps $\Delta E_\mathrm{HL}$, grand-canonical frontier orbital spacings $\Delta_\mu$, and effective broadening scales $k_\mathrm{B}T_\mathrm{eff}$ extracted from the fractional occupations; error bars on $k_\mathrm{B}T_\mathrm{eff}$ denote the standard error of the linear fit used to extract the effective temperature (see SI Section S3.4). Lower panel: ratio $k_\mathrm{B}T_\mathrm{eff}/\Delta_\mu$ for the optimized fragments. Values are reported in Table~S6.
\label{fig:hring_convergence}
}
\end{figure}

These observations motivate a systematic analysis of the fragment-size dependence. In the following, we therefore reoptimize the self-energy parameters for a series of fragments from \ce{H6} to \ce{H20} and show in Fig.~\ref{fig:hring_convergence}a how the optimized coupling parameters converge. The resulting potential curves are shown in Fig.~S11 in the SI.
\setlength{\textfloatsep}{3pt}
The optimized coupling parameters exhibit a systematic trend with fragment size (see Fig.~\ref{fig:hring_convergence}). Both the orbital-energy shift parameter $\varepsilon$ and the broadening parameter $\eta$ decrease with increasing fragment size and approach zero as a larger fraction of the environment is treated explicitly. This trend directly reflects the diminishing need for an effective reservoir: as more of the quasi-periodic system is described explicitly, progressively less self-energy correction is required to reproduce the reference response, and in the limit of the full periodic system, no reservoir is needed at all, so the self-energy correction must vanish. The observed convergence demonstrates that the self-energy acts as a compact surrogate for the missing environment.

Density-difference plots for the optimized fragments are provided in the SI (see Fig.~S12). The detailed spatial pattern of the density deviations depends on fragment size and does not show a strictly monotonic local trend, most likely due to concurring effects such as saturation of the undercoordinated ends and compensation for \acp{MO} spread out over the whole system. Nevertheless, the largest deviations are observed for the smallest fragments, consistent with the larger coupling parameters required there.

To further characterize the electronic structure underlying the metal-like response, we computed HOMO--LUMO gaps, grand-canonical frontier orbital spacings, and effective temperatures (see SI Section S3.4 for details) derived from the fractional occupations, as summarized in Fig.~\ref{fig:hring_convergence}b and Table~S6. Since the frontier orbitals acquire fractional occupations in the grand-canonical framework, the conventional HOMO--LUMO distinction becomes less meaningful. We therefore additionally define a grand-canonical frontier orbital spacing, $\Delta_\mu$, as the energy difference between the closest molecular orbitals below and above the chemical potential.

The chemical potentials obtained from the neutrality constraint, together with the canonical and grand-canonical frontier levels, are shown for all fragment sizes in the SI (see Fig.~S13). For all fragments, the optimized chemical potential lies closer to the HOMO than to the canonical midgap value, though this asymmetry narrows with increasing fragment size as $\mu$ gradually approaches the midgap value. This trend is consistent with finite-size effects in the smaller fragments, which would otherwise accumulate excess electronic charge under the grand-canonical broadening to saturate the edges; as a larger fraction of the quasi-periodic system is treated explicitly, progressively smaller shifts of the chemical potential are required to maintain charge neutrality. Correspondingly, the grand-canonical HOMO level converges onto the canonical HOMO from \ce{H14} onward, essentially coinciding with it by \ce{H18}/\ce{H20}, mirroring the vanishing self-energy correction seen in $\varepsilon$ and $\eta$.

While the \ce{H6} fragment retains a molecular HOMO--LUMO gap of approximately 0.38~Ha, the full \ce{H40} system exhibits quasi-metallic behavior with a substantially smaller gap of only 0.04~Ha. Interestingly, the nearly periodic \ce{H38} ring still exhibits a HOMO--LUMO gap of 0.11~Ha, substantially larger than the value obtained for \ce{H40}, despite differing by only two hydrogen atoms. This pronounced increase in the gap is consistent with the reduced polarization response observed for \ce{H38} and illustrates the sensitivity of the electronic structure to the presence or absence of the periodicity. For the optimized fragments, the grand-canonical frontier orbital spacings $\Delta_\mu$ are consistently smaller than the corresponding canonical HOMO--LUMO gaps. However, these spacings remain substantially larger than the HOMO--LUMO gap of the \ce{H40} reference system. 
The metal-like charge response of the fragment, therefore, does not arise from a complete collapse of the electronic gap. Instead, it results primarily from the occupation broadening and fractional occupations introduced by the grand-canonical treatment.

An effective electronic temperature extracted from the fractional occupations yields $T_\mathrm{eff}\sim 2$--$3\times10^4$ K, quantifying the width of the occupation transition around the chemical potential. As shown in Fig.~\ref{fig:hring_convergence}b, the effective broadening scale $k_\mathrm{B}T_\mathrm{eff}$ remains a rather constant fraction of the grand-canonical frontier orbital spacing $\Delta_\mu$ throughout the fragment series, despite the pronounced variation in the underlying orbital spectrum.
This observation suggests that the optimized self-energy adapts to the intrinsic electronic structure of the fragment, providing sufficient occupation broadening to reproduce the metallic-like polarization response of the full ring. These results emphasize the importance of allowing fractional occupations when modeling metallic systems using finite-cluster models.

Overall, the hydrogen-ring model demonstrates that a local grand-canonical self-energy can recover the response of an extended quasi-metallic system, provided that the coupling parameters are optimized for the finite fragment. The optimized self-energy compensates for the missing environment by enabling fractional occupations and charge redistribution, while its systematic decrease with fragment size confirms that this correction vanishes as the explicit quantum region approaches the reference system. The remaining deviations can therefore be attributed primarily to finite-size truncation of the polarization response rather than to a failure of the grand-canonical formulation itself.

\subsection{Lithium cluster}
\begin{figure}
\includegraphics[width=\linewidth]{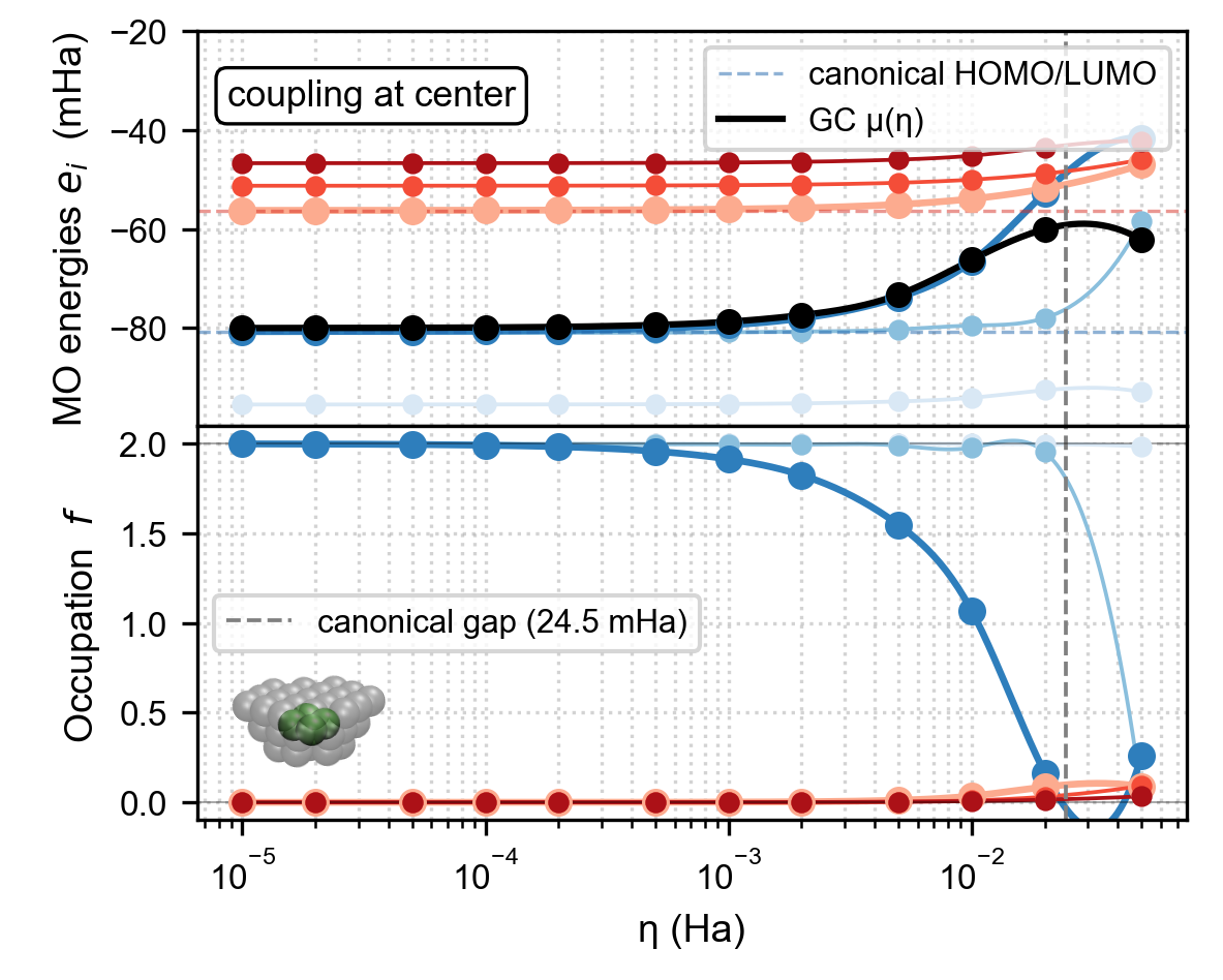}
\caption{
\ac{MO} energies $e_i$ (upper panel) and
grand-canonical occupations
$f_i= \mathbf{C}_i^\top \mathbf{S}\,\mathbf{P}\,\mathbf{S}\,\mathbf{C}_i$
(lower panel) as a function of the broadening $\eta$ for the
\ce{Li40} cluster coupled at the four central
bulk atoms, where $\mathbf{P}(\eta)$ is the converged grand-canonical density
matrix, $\mathbf{S}$ the atomic-orbital overlap matrix, and $\mathbf{C}_i$ the
fixed canonical ($\eta=0$) molecular-orbital coefficients of orbital $i$.
Blue curves show the three highest occupied orbitals
and red curves the
three lowest unoccupied orbitals.
Dashed horizontal lines mark the canonical \ac{HOMO} and \ac{LUMO}
energies where the self-consistent chemical potential $\mu(\eta)$ is
shown in black. The vertical dashed line indicates the canonical
\ac{HOMO}--\ac{LUMO} gap (24.5~mHa).
\label{fig:broadening_center}
}

\end{figure}

\vspace{-0.5cm}
As a final test case, we study a \ce{Li40} cluster. We first investigate the effect of complex-energy broadening $\eta$ on the orbital spectrum and occupations without any orbital shift ($\varepsilon=0$). To reduce edge effects, we apply the coupling to the four center atoms, which are enclosed by the outer atoms. The chemical potential is self-consistently adjusted at each broadening value to achieve charge neutrality. The results are shown in Fig.~\ref{fig:broadening_center}.

\begin{figure}[h!]
\includegraphics[width=\linewidth]{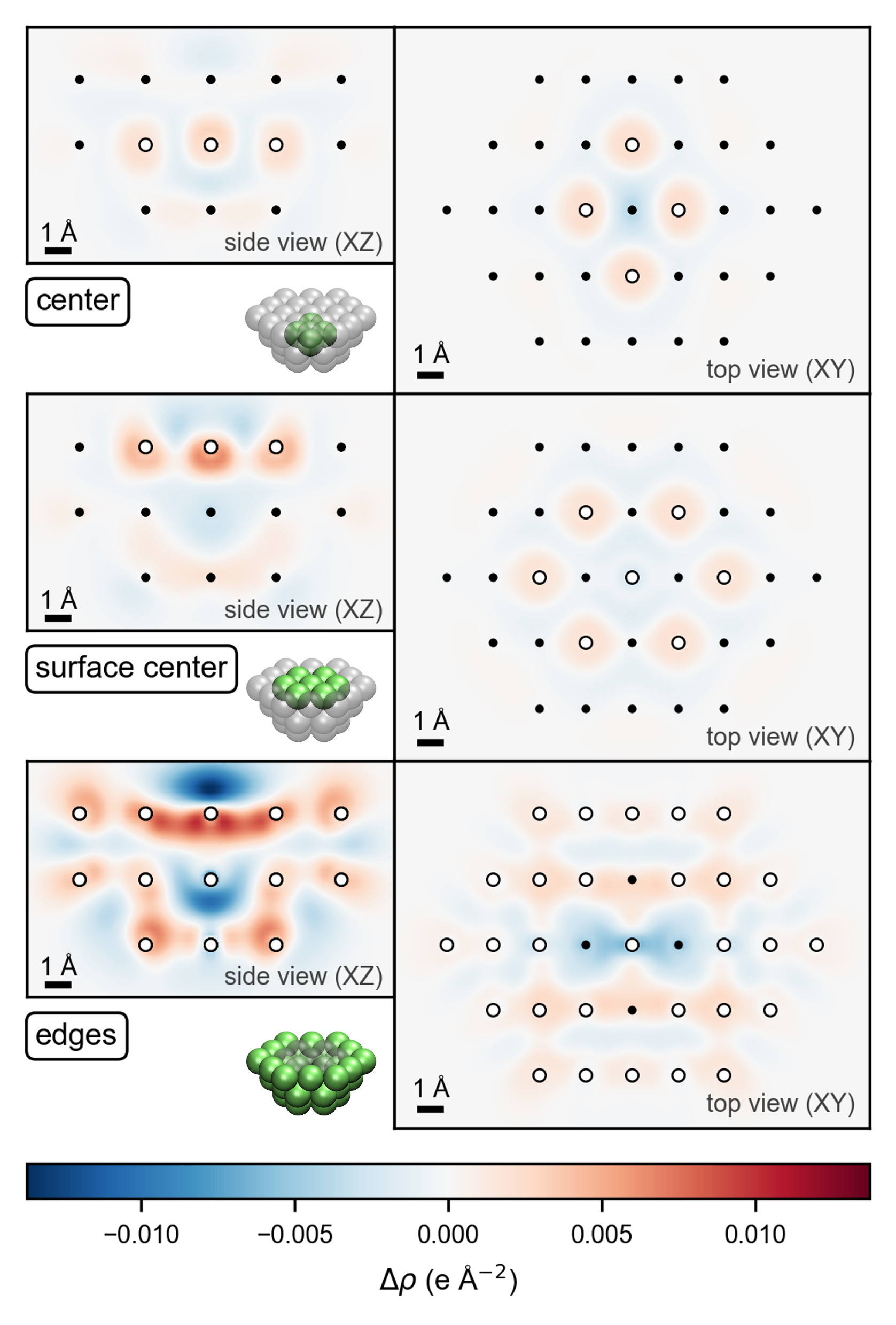}
\caption{
Density difference $\Delta\rho(\mathbf{r}) = \rho^\mathrm{GC}(\mathbf{r}) - \rho^\mathrm{can}(\mathbf{r})$ for the Li$_{40}$ cluster at broadening parameter $\eta = 5$ mHa. Results are shown for three coupling topologies: center (left), surface center (middle), and edges (right), and two different projections. The upper row shows the density integrated along the $y$-axis and plotted in the $xz$-plane (side view), while the lower row shows it integrated along the $z$-axis and plotted in the $xy$-plane (top view), both in e\,\AA$^{-2}$. Black circles indicate all cluster atoms, while white circles mark the coupled atoms. All calculations were carried out at the PBE0/STO-3G level.
\label{fig:li_density}}
\end{figure}
For $\eta\rightarrow 0$ the cluster recovers the canonical orbital spectrum. The chemical potential remains pinned close to the HOMO energy rather than at the middle of the HOMO–LUMO gap, reflecting the tendency of the undercoordinated surface Li atoms to accept electrons from the reservoir. This illustrates that finite metal clusters alone cannot fully reproduce the electronic structure of an extended electrode and therefore motivates combining the present open-boundary formalism with complementary embedding approaches.\cite{lee_bath_2026}
As $\eta$ approaches the magnitude of the canonical \ac{HOMO}--\ac{LUMO} gap of $24.5~\text{mHa}$, the broadening simultaneously samples both frontier orbitals and the occupation of the \ac{HOMO} drops rapidly, falling from $1.55$ at $\eta =5~\text{mHa}$ to $1.07$ at $\eta = 10~\text{mHa}$ and nearly vanishing at $\eta =20~\text{mHa}$. In the latter case, charge neutrality is achieved by a finite, low occupation of a large number of previously unoccupied orbitals.
Concurrently, the eigenvalues shift substantially to higher values and the effective gap collapses. This defines a natural, physically meaningful regime $\eta \ll \Delta\varepsilon_\mathrm{gap}$, beyond which the broadening distorts the electronic structure rather than modeling the open-boundary coupling.

In the following, we analyze how the coupling geometry affects the induced density redistribution. Therefore, we apply the coupling to three different sets of atoms: the center coupling we already introduced, edge coupling, and surface center coupling. The locations of the coupled atoms are indicated in green in Fig.~\ref{fig:li_density}. The density difference $\Delta\rho$ is small throughout but most pronounced for the edge coupling scheme, which involves the largest number of coupled atoms. This is consistent with the approximately linear scaling of the integrated density perturbation with the number of coupled atoms (Fig.~S16), and mirrors the trend already visible in the broadening dependent orbital spectrum (see Fig.~S15), where the eigenvalue shifts at a given $\eta$ are largest for the edges scheme. 

\begin{figure}
\includegraphics[width=0.95\linewidth]{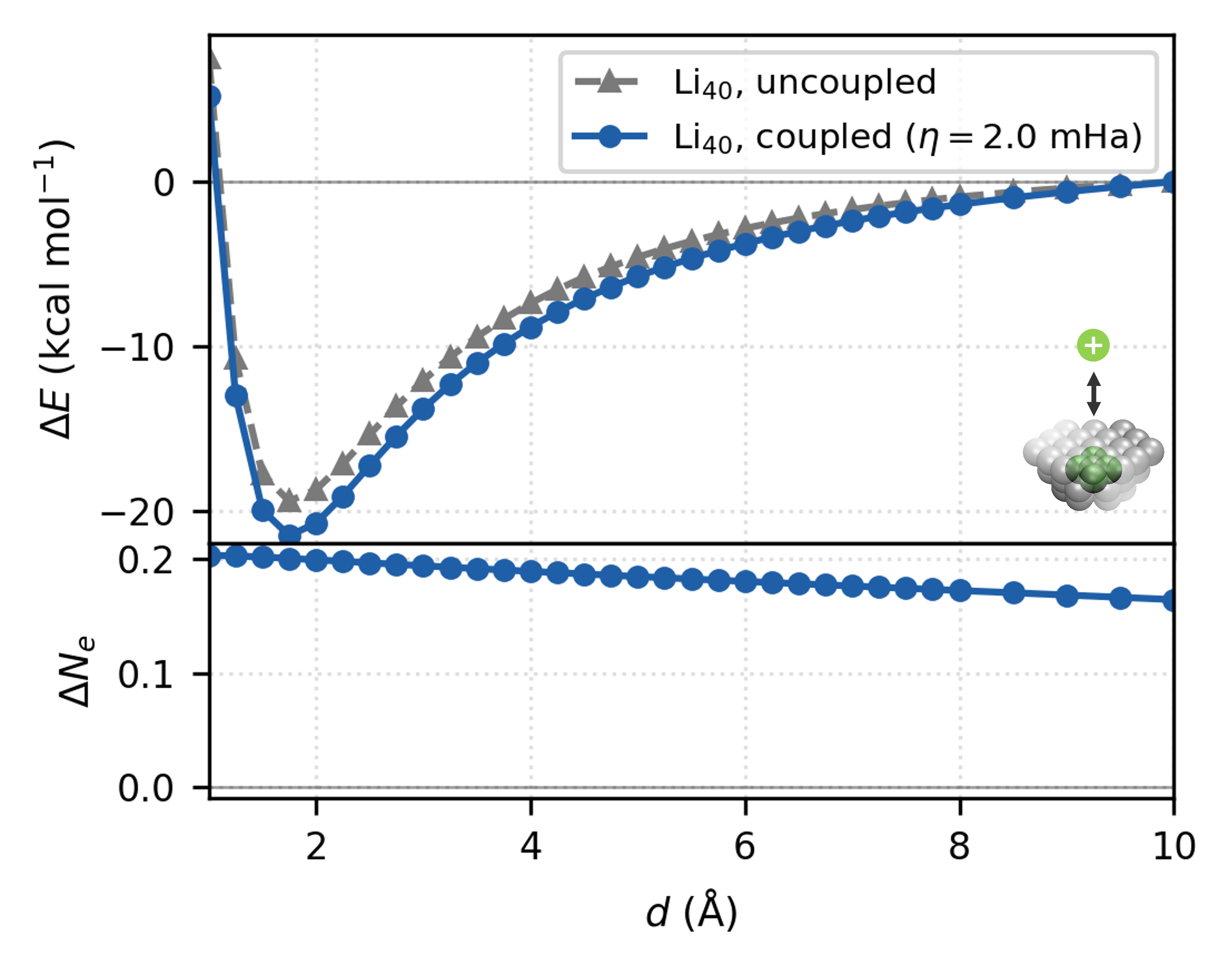}
\caption{
Potential energy curve $\Delta E(d)$ (upper panel) and change in electron count
$\Delta N_e(d)$ (lower panel) as a function of the approach distance $d$ of a
positive point charge $q = + 0.4\,e$ above the \ce{Li40} cluster surface,
computed with the center coupling scheme ($\varepsilon = 0$,
$\eta = 2.0~\text{mHa}$, PBE0/STO-3G).
Energies are reported relative to $d = 10~\text{\AA}$.
The \ac{GC-SCF} result (blue) is compared to a canonical 
reference calculation (grey).
\label{fig:charge_scan}
}
\end{figure}

To illustrate the ability of the \ac{GC-SCF} framework to model open-boundary charge transfer, we calculated a similar potential curve as in the previous section, where a point-charge approaches the cluster surface using the center coupling scheme. The latter has been chosen because it introduces the smallest perturbation to the electronic density among the three coupling topologies considered. A positive point charge $q~=~+0.4\,e$ is placed along the surface normal above a surface atom and moved towards the cluster. The chemical potential is fixed at the value that retains charge neutrality of the unperturbed cluster, so that the electron count responds freely as the external charge approaches. The results, along with a canonical reference calculation, are shown in Fig.~\ref{fig:charge_scan}. The \ac{GC-SCF} adsorption potential exhibits a well of approximately $21.5~\text{kcal/mol}$ at $d \approx 1.75~\text{\AA}$ relative to the large-distance limit, modestly deeper than the canonical reference ($19.4~\text{kcal/mol}$), which arises from polarization alone.
The energetic difference between the two curves is indicative of the stabilization due to charge accumulation as a response to this external perturbation. Concurrently, $\Delta N_e$ increases monotonically up to $+0.20\,e$ at $d = 1~\text{\AA}$, reflecting electron inflow from the reservoir in response to the approaching positive charge.
Taken together, the lithium-cluster calculations demonstrate that the grand-canonical boundary conditions in the \ac{GC-SCF} framework enable a metal-like charge-polarization response in finite clusters via localized coupling to the external reservoir.

\section{Conclusion}

We have presented a grand-canonical self-consistent field framework that enables open-boundary electronic-structure calculations for finite quantum-chemical clusters by incorporating an energy-independent complex self-energy into conventional Hartree–Fock and density functional theory. Within the wide-band approximation, the resulting working equations admit an analytic solution and can be integrated into existing electronic-structure methods with only minor modifications.

The numerical robustness of the framework was demonstrated for a single water molecule, where the non-Hermitian self-consistent field equations converge reliably using standard Pulay-type mixing schemes. For a quasi-periodic hydrogen ring, we showed that a finite fragment coupled to an optimized self-energy accurately reproduces the polarization response of the extended system. The systematic decrease of the optimized coupling parameters with increasing fragment size demonstrates that the self-energy acts as a compact representation of the missing environment rather than introducing an artificial electronic response. Finally, calculations on a \ce{Li40} cluster show that the proposed boundary conditions reproduce the characteristic metallic charge response under open-boundary conditions through a localized coupling to the external reservoir.

These results establish a practical set of grand-canonical boundary conditions for finite quantum-chemical clusters that capture essential open-system effects such as fractional occupations, charge transfer, and metallic screening. Rather than constituting a complete embedding framework, the present work lays the methodological foundation for future grand-canonical quantum embedding methods applicable to electrochemical systems.

Future work will therefore focus on combining the present open-boundary formalism with other quantum embedding techniques and solvation models to enable the application in realistic electrochemical interfaces. A special focus will lie on an automated and transferable parameterization for the self-energy. This will enable finite cluster models to exchange electrons with an external reservoir while treating the chemically active region at an appropriate electronic-structure level. Moreover, motivated by recent advances in grand-canonical correlated electronic-structure methods,\cite{wei2022improving,cheng2026quantifying} future work will investigate extending the present framework beyond the mean-field level to incorporate electronic correlation through the random phase approximation (RPA).

\begin{acknowledgments}
This publication is part of the HeliECat project that has received funding from the European Research Council (ERC) under the European Union’s Horizon research and innovation programme (Grant agreement No. 101219789).
Funded by the European Union. Views and opinions expressed are however those of the author(s) only and do not necessarily reflect those of the European Union or the European Research Council Executive Agency. Neither the European Union nor the granting authority can be held responsible for them.
Early stages of the project were funded through an Early Postdoc.Mobility fellowship from the Swiss National Science Foundation. 
MHG acknowledges support from the Liquid Sunlight Alliance, which is
funded by the U.S. Department of Energy, Office of Science, Office
of Basic Energy Sciences, Fuels from Sunlight Hub under Award
Number DE-SC0021266.
L.H. gratefully acknowledges financial support from the Fonds der Chemischen Industrie through a Kekulé Fellowship during the initial phase of this project. L.H. further acknowledges support from the Studienstiftung des deutschen Volkes through a doctoral fellowship and from the Marianne-Plehn-Program of the Elite Network of Bavaria, funded by the Free State of Bavaria. 
Finally, L.H. thanks Elena Kolodzeiski for insightful discussions on the metal cluster calculations.

\end{acknowledgments}

\section*{Data Availability Statement}

The modifications to pySCF implemented to obtain the presented results will be made available upon acceptance of the article in a peer-reviewed journal.


\appendix

\section{Derivation of Eq.~\ref{eq:P_Gimg}}\label{app:eq:P_Gimg}

We impose an orthonormal basis such that the overlap matrix reduces to
$\mathbf{S}=\mathbf{I}$. Assuming a Hermitian Hamiltonian matrix $\mathbf{H}$ and real energies $E$, one obtains,
following Datta~\cite{datta2000nanoscale},
\begin{eqnarray}
    \mathbf{G}_R^{-1}-(\mathbf{G}_R^\dagger)^{-1}
    &=&  E\mathbf{I}-\mathbf{H}-\boldsymbol{\Sigma}
    -E\mathbf{I}+\mathbf{H}^\dagger+\boldsymbol{\Sigma}^\dagger \notag\\
    &=& \boldsymbol{\Sigma}^\dagger-\boldsymbol{\Sigma} = - \frac{\boldsymbol{\Gamma}}{i} = i\boldsymbol{\Gamma} .
\end{eqnarray}
Multiplying from the left by $\mathbf{G}_R$ and from the right by
$\mathbf{G}_R^\dagger$ yields
\begin{eqnarray}\label{eq:G_diff}
    \mathbf{G}_R^\dagger-\mathbf{G}_R
    &=&
    \mathbf{G}_R\underbrace{(\boldsymbol{\Sigma}^\dagger-\boldsymbol{\Sigma})}_{i\boldsymbol{\Gamma}}
    \mathbf{G}_R^\dagger   \notag\\ &=&
    i\,\mathbf{G}_R\boldsymbol{\Gamma}\mathbf{G}_R^\dagger .
\end{eqnarray}
where the definition of the energy broadening (see Eq. \ref{eq:gamma}) was used. Hence, we arrive at the following equality

\begin{equation}
    \mathbf{G}_R\boldsymbol{\Gamma}\mathbf{G}_R^\dagger = -i\left(  \mathbf{G}_R^\dagger-\mathbf{G}_R \right) = i\left(  \mathbf{G}_R-\mathbf{G}_R^\dagger \right).
\end{equation}

\section{Derivation of Eq. \ref{eq:density_final}}\label{app:deriv_final}

Combining Eq.~\ref{eq:Gr} and Eq.~\ref{eq:B_eigenvalue} yields
\begin{eqnarray}
    \mathbf{G}_R^{-1} &=& E\mathbf{S}-\mathbf{H}_\text{eff} \notag\\
    \Leftrightarrow\quad
    \mathbf{G}_R^{-1} &=& E\mathbf{S}-\mathbf{B}\mathbf{Z}\mathbf{B}^{-1}.
\end{eqnarray}
We again impose an orthonormal basis such that the overlap matrix reduces to
$\mathbf{S}=\mathbf{I}$, which gives
\begin{eqnarray}
    \mathbf{G}_R
    &=&
    \left(E\mathbf{I}-\mathbf{B}\mathbf{Z}\mathbf{B}^{-1}\right)^{-1} \notag\\
    &=&
    \mathbf{B}\left(E\mathbf{I}-\mathbf{Z}\right)^{-1}\mathbf{B}^{-1} \label{eq:arnold26}
\end{eqnarray}
as a result of the above similarity transformation. The Hermitian conjugate of the retarded Green's function is given by
\begin{eqnarray}
    \mathbf{G}_R^\dagger
    &=&
    \left[
    \mathbf{B}
    \left(E\mathbf{I}-\mathbf{Z}\right)^{-1}
    \mathbf{B}^{-1}
    \right]^\dagger
    \notag\\
    &=&
    \left(\mathbf{B}^{-1}\right)^\dagger
    \left(E\mathbf{I}-\mathbf{Z}^*\right)^{-1}
    \mathbf{B}^\dagger .
\end{eqnarray}
where we used that $\mathbf{Z}$ is diagonal and $E$ is real.\\
Using Eq.~\ref{eq:P_Gimg}, the density matrix can therefore be written as
\begin{eqnarray}
    \mathbf{P}
    &=&
    \frac{i}{2\pi}
    \int_{-\infty}^{\mu}
    \mathrm{d}E\,
    \Bigg[
    \mathbf{B}
    \left(E\mathbf{I}-\mathbf{Z}\right)^{-1}
    \mathbf{B}^{-1}
    \notag\\
    &&\qquad\qquad -
    \left(\mathbf{B}^{-1}\right)^\dagger
    \left(E\mathbf{I}-\mathbf{Z}^*\right)^{-1}
    \mathbf{B}^{\dagger}
    \Bigg].
\end{eqnarray}
Because the coupled orbitals carry a strictly positive broadening ($\eta_{pp} > 0$), every eigenvalue $Z_k$ has a strictly negative imaginary part. The integrand therefore has no pole on the real axis, and $E\mathbf{I} - \mathbf{Z}$ stays in the upper half-plane along the entire contour, so the principal-branch logarithm is a valid antiderivative.
The above integral evaluates to
\begin{eqnarray}
\mathbf{P}
&&= \frac{i}{2\pi}
\Biggl[
\mathbf{B}\Bigl(
\ln(\mu\mathbf{I}-\mathbf{Z})
-\lim_{E\to-\infty}\ln(E\mathbf{I}-\mathbf{Z})
\Bigr)\mathbf{B}^{-1}
\notag\\
&&-
(\mathbf{B}^{-1})^\dagger
\Bigl(
\ln(\mu\mathbf{I}-\mathbf{Z}^*)
-\lim_{E\to-\infty}\ln(E\mathbf{I}-\mathbf{Z}^*)
\Bigr)\mathbf{B}^\dagger
\Biggr].
\label{eq:P_logs_with_limits}
\end{eqnarray}
Using the relation
\begin{equation}
    \ln z = \ln |z| + i\,\arg(z),
\end{equation}
the contributions from the lower integration limit can be written as
\begin{eqnarray}
\ln(E\mathbf{I}-\mathbf{Z})
&=&
\ln|E\mathbf{I}-\mathbf{Z}|
+i\,\arg(E\mathbf{I}-\mathbf{Z}),
\notag\\
\ln(E\mathbf{I}-\mathbf{Z}^*)
&=&
\ln|E\mathbf{I}-\mathbf{Z}^*|
+i\,\arg(E\mathbf{I}-\mathbf{Z}^*).
\end{eqnarray}
Since the self-energy has a negative-semidefinite imaginary part ($\Sigma_{pp} = \varepsilon_{pp} - i \eta_{pp}$ and $\eta >0$),
the eigenvalues of $\mathbf{H}_{\mathrm{eff}}$ lie in the lower half of
the complex plane. Consequently, on the principal branch,
\begin{eqnarray}
\lim_{E\to-\infty}
\arg(E\mathbf{I}-\mathbf{Z})
&=&
\pi\mathbf{I},
\notag\\
\lim_{E\to-\infty}
\arg(E\mathbf{I}-\mathbf{Z}^*)
&=&
-\pi\mathbf{I}.
\end{eqnarray}
Furthermore,
\begin{equation}
\lim_{E\to-\infty}
\left[
\ln|E\mathbf{I}-\mathbf{Z}|
-
\ln|E\mathbf{I}-\mathbf{Z}^*|
\right]
=0.
\end{equation}
Hence,
\begin{equation}
\lim_{E\to-\infty}
\left[
\ln(E\mathbf{I}-\mathbf{Z})
-
\ln(E\mathbf{I}-\mathbf{Z}^*)
\right]
=
2\pi i\,\mathbf{I}.
\end{equation}

In the limit $E \rightarrow -\infty$, each logarithm becomes proportional to the identity, $\ln(E\mathbf{I} - \mathbf{Z}) \rightarrow \ln|E| \mathbf{I} + i\pi\mathbf{I}$ and $\ln(E\mathbf{I} - \mathbf{Z}^*) \rightarrow \ln|E| \mathbf{I} - i\pi \mathbf{I}$, so both similarity transformations in Eq. B5 act trivially. 
The divergent $\ln|E| \mathbf{I}$ contributions are identical [cf. Eq. B9] and cancel between the two terms, while the remaining $\pm i\pi\mathbf{I}$ parts combine. 
Hence, the lower integration limit contributes $-2\pi i\mathbf{I}$ to Eq. B5, and the density matrix becomes

\begin{eqnarray}
\mathbf{P}
&=&
\mathbf{I}
+\frac{i}{2\pi}
\Bigl[
\mathbf{B}\ln(\mu\mathbf{I}-\mathbf{Z})\mathbf{B}^{-1}
\notag\\
&&-
(\mathbf{B}^{-1})^\dagger
\ln(\mu\mathbf{I}-\mathbf{Z}^*)
\mathbf{B}^\dagger
\Bigr].
\label{eq:P_logs_eigbasis}
\end{eqnarray}
To obtain a more compact expression, we shift the arguments of the
complex logarithms. Since the eigenvalues of the retarded effective
Hamiltonian lie in the lower half of the complex plane, the principal
branch gives
\begin{eqnarray}
\ln(\mu\mathbf{I}-\mathbf{Z})
&=&
\ln(\mathbf{Z}-\mu\mathbf{I})
+i\pi\mathbf{I},
\notag\\
\ln(\mu\mathbf{I}-\mathbf{Z}^*)
&=&
\ln(\mathbf{Z}^*-\mu\mathbf{I})
-i\pi\mathbf{I}.
\end{eqnarray}
Substitution into Eq.~\ref{eq:P_logs_eigbasis} yields
\begin{eqnarray}
\mathbf{P}
&=&
\mathbf{I}
+\frac{i}{2\pi}
\Bigl[
\mathbf{B}\ln(\mathbf{Z}-\mu\mathbf{I})\mathbf{B}^{-1}
\notag\\
&&-
(\mathbf{B}^{-1})^\dagger
\ln(\mathbf{Z}^*-\mu\mathbf{I})
\mathbf{B}^\dagger
+2\pi i\,\mathbf{I}
\Bigr].
\end{eqnarray}
The constant contribution cancels the explicit identity matrix,
such that
\begin{eqnarray}
\mathbf{P}
&=&
\frac{i}{2\pi}
\Bigl[
\mathbf{B}\ln(\mathbf{Z}-\mu\mathbf{I})\mathbf{B}^{-1}
\notag\\
&&-
(\mathbf{B}^{-1})^\dagger
\ln(\mathbf{Z}^*-\mu\mathbf{I})
\mathbf{B}^\dagger
\Bigr].
\label{eq:P_logs_shifted}
\end{eqnarray}

Finally, using
$\mathbf{H}_{\mathrm{eff}}=\mathbf{B}\mathbf{Z}\mathbf{B}^{-1}$
and the invariance of matrix functions under similarity transformations,
we obtain
\begin{equation}
\mathbf{P}
=
\frac{i}{2\pi}
\left[
\ln(\mathbf{H}_{\mathrm{eff}}-\mu\mathbf{I})
-
\ln(\mathbf{H}_{\mathrm{eff}}^\dagger-\mu\mathbf{I})
\right].
\label{eq:P_log_Heff}
\end{equation}

Orbitals not coupled to the reservoir ($\eta_{pp} = 0$) yield eigenvalues on the real axis; these are recovered in the limit $\eta \rightarrow 0^+$, which reproduces their integer occupations.

\nocite{*}
\bibliography{refs}

\end{document}


\newpage
\tableofcontents
\clearpage
\section{S1 Details of SCF Mixing Schemes}
\addcontentsline{toc}{section}{S1 Details of SCF Mixing Schemes}
Here, we report implementation details for the implemented SCF mixing schemes. 

\paragraph{CDIIS.}
In the CDIIS scheme, the Fock matrix $\mathbf{F}$ is mixed by minimizing the norm of the commutator residual
\begin{equation}
\mathbf{R} = \mathbf{F}\mathbf{P}\mathbf{S} - \mathbf{S}\mathbf{P}\mathbf{F}.
\end{equation}
The Pulay $B$ matrix is constructed from scalar products of residual vectors.
If the linear system becomes ill-conditioned, a diagonal regularization term of $10^{-12}$ is added.
The CDIIS implementation closely follows the corresponding implementation in PySCF version~2.12.0, adapted to the present grand-canonical SCF framework.

\paragraph{EDIIS(E).}
In the energy-DIIS variant, Fock matrices are combined by minimizing the total-energy functional (see, e.g., Ref.~\citenum{garza2012comparison})
\begin{equation}
E_{\mathrm{tot}}[\mathbf{P}] = E_{\mathrm{elec}}[\mathbf{P}] + E_{\mathrm{nuc}},
\end{equation}
subject to convexity constraints on the mixing coefficients.

\paragraph{EDIIS($\Omega$).}
In the grand-canonical formulation, the Pulay functional is constructed from the grand potential
\begin{equation}
\Omega[\mathbf{P}] = E_{\mathrm{tot}}[\mathbf{P}] 
- \mu \, \mathrm{Tr}[\mathbf{P}\mathbf{S}],
\end{equation}
which replaces the total energy as the target functional in the EDIIS optimization.

\paragraph{Anderson mixing.}
Anderson acceleration was applied directly to the density matrix.
In each SCF iteration, a density matrix $\mathbf{P}_{\mathrm{in}}$ is used to construct the Fock matrix and, after diagonalization of the effective Hamiltonian and occupation update, a new density matrix $\mathbf{P}_{\mathrm{out}}$ is obtained. Accordingly, the residual used in Anderson acceleration is defined as
\begin{equation}
\mathbf{f} =
\mathbf{P}_{\mathrm{out}} - \mathbf{P}_{\mathrm{in}}.
\end{equation}
For standalone Anderson mixing, acceleration was activated after two initial iterations, during which simple linear mixing was employed.
The Anderson update was obtained from a regularized least-squares problem using residual and density differences from the stored history.
A damping factor of 0.1 was used for standalone Anderson mixing, whereas a damping factor of 0.3 was employed in the hybrid EDIIS($\Omega$)+A scheme (see below).
If the coefficient norm exceeded the predefined threshold of 10, the accelerated step was discarded and linear mixing was used instead.

\paragraph{Hybrid EDIIS($\Omega$)+A.}
The hybrid scheme employs EDIIS($\Omega$) during the initial phase of the SCF procedure to ensure global stability.
After at least 12 SCF iterations, and once the density change satisfied $\|\Delta \mathbf{P}\| < 10^{-2}$, the algorithm switches to Anderson acceleration.
During the Anderson phase, the same density-matrix Anderson update as in the standalone Anderson scheme was used.
To improve numerical stability, accelerated Anderson steps were rejected whenever the grand potential of the accelerated density exceeded that of the corresponding unaccelerated fixed-point step.
In such cases, the unaccelerated trial density was accepted instead, and the Anderson history was reset.

\newpage
\section{S2 Further results and data from convergence tests with water}
\addcontentsline{toc}{section}{S2 Further results and data from convergence tests with water}
In the following subsections, we report the results of the SCF convergence tests on a single water molecule (see \ref{fig:h2o_xyz} for coordinates) using different levels of theory, basis sets, and self-energies. For each level of theory and basis set combination, we report results for three values of $\eta$, using different mixing schemes that start from two different initial guesses for the density matrix.
\begin{figure}[H]
\centering

\begin{tcolorbox}[
  colback=gray!10,
  colframe=gray!40,
  boxrule=0.3pt,
  arc=2pt,
  width=\textwidth
]
\begin{lstlisting}
3

O  0.0000000   0.0000000   0.0000000
H  0.0000000  -0.7570000   0.5870000
H  0.0000000   0.7570000   0.5870000

\end{lstlisting}
\end{tcolorbox}

\caption{Cartesian coordinates (in Å) of the \ce{H2O} molecule used in this work.}
\label{fig:h2o_xyz}

\end{figure}
\newpage
\subsection{S2.1 Restricted Hartree-Fock}
\addcontentsline{toc}{subsection}{S2.1 Restricted Hartree-Fock}
\begin{table}
\caption{Final grand-canonical SCF results for the water molecule using the \textbf{STO-3G} basis at \textbf{RHF} level. Reported are the grand potential $\Omega$, the number of electrons $N_e$, and the Frobenius norm of the FPS residual $\|\mathbf{R}\|_F$. Residuals are reported only for converged calculations; $\dagger$ marks a run that did not meet the convergence criterion. The reference chemical potential is $\mu_{\mathrm{ref}} = 0.10693~\mathrm{Ha}$.}
\label{tab:water_scf_results_log}
\centering
\begin{tabular}{l|cccc|cccc}
\hline
& \multicolumn{4}{c|}{SAD guess} & \multicolumn{4}{c}{$\boldsymbol{\Sigma}=0$ start} \\
SCF scheme & $\Omega$ (Ha) & $N_e$ & iter & $\|\mathbf{R}\|_F$ & $\Omega$ (Ha) & $N_e$ & iter & $\|\mathbf{R}\|_F$ \\
\hline
\multicolumn{9}{c}{$\eta = 1.0\times10^{-6}$ Ha} \\
\hline
EDIIS(E) & -76.03237 & 10.00000 & 59 & $3.15\times10^{-6}$ & -76.03237 & 10.00000 & 3 & $1.54\times10^{-6}$ \\
CDIIS & -76.03237 & 10.00000 & 9 & $1.43\times10^{-6}$ & -76.03237 & 10.00000 & 4 & $1.45\times10^{-6}$ \\
Anderson & -76.03237 & 10.00000 & 13 & $1.45\times10^{-6}$ & -76.03237 & 10.00000 & 40 & $1.45\times10^{-6}$ \\
EDIIS($\Omega$) & -76.03237 & 10.00000 & 59 & $3.15\times10^{-6}$ & -76.03237 & 10.00000 & 3 & $1.54\times10^{-6}$ \\
EDIIS($\Omega$)+A & -76.03237 & 10.00000 & 22 & $1.46\times10^{-6}$ & -76.03237 & 10.00000 & 3 & $1.54\times10^{-6}$ \\
\hline
\multicolumn{9}{c}{$\eta = 5\times10^{-3}$ Ha} \\
\hline
EDIIS(E) & -76.00512 & 10.01086 & 59 & $7.15\times10^{-3}$ & -76.00512 & 10.01086 & 33 & $7.15\times10^{-3}$ \\
CDIIS & -75.97226$^\dagger$ & 10.01124 & 201 & -- & -76.00518 & 10.01112 & 3 & $7.41\times10^{-3}$ \\
Anderson & -76.00512 & 10.01086 & 12 & $7.15\times10^{-3}$ & -76.00512 & 10.01086 & 87 & $7.15\times10^{-3}$ \\
EDIIS($\Omega$) & -76.00512 & 10.01086 & 59 & $7.15\times10^{-3}$ & -76.00512 & 10.01086 & 33 & $7.15\times10^{-3}$ \\
EDIIS($\Omega$)+A & -76.00512 & 10.01086 & 22 & $7.15\times10^{-3}$ & -76.00512 & 10.01086 & 31 & $7.15\times10^{-3}$ \\
\hline
\multicolumn{9}{c}{$\eta = 1\times10^{-1}$ Ha} \\
\hline
EDIIS(E) & -75.49681 & 10.13894 & 72 & $1.19\times10^{-1}$ & -75.49681 & 10.13894 & 65 & $1.19\times10^{-1}$ \\
CDIIS & -53.06557$^\dagger$ & 2.30500 & 201 & -- & -75.52359 & 10.19133 & 10 & $1.24\times10^{-1}$ \\
Anderson & -75.49681 & 10.13894 & 61 & $1.19\times10^{-1}$ & -75.49681 & 10.13894 & 123 & $1.19\times10^{-1}$ \\
EDIIS($\Omega$) & -75.49681 & 10.13894 & 72 & $1.19\times10^{-1}$ & -75.49681 & 10.13894 & 65 & $1.19\times10^{-1}$ \\
EDIIS($\Omega$)+A & -75.49681 & 10.13894 & 27 & $1.19\times10^{-1}$ & -75.49681 & 10.13894 & 43 & $1.19\times10^{-1}$ \\
\hline
\end{tabular}
\end{table}

\begin{figure}[H]
    \centering

    \begin{minipage}{0.49\textwidth}
        \centering
        \includegraphics[width=\linewidth]
        {Figures/SI Figures/water_convergence__rhf__basis-sto-3g__start-canonical_same_theory__density-log.png}
    \end{minipage}
    \hfill
    \begin{minipage}{0.49\textwidth}
        \centering
        \includegraphics[width=\linewidth]
        {Figures/SI Figures/water_convergence__rhf__basis-sto-3g__start-sad__density-log.png}
    \end{minipage}

    \caption{
    SCF convergence behavior for the grand-canonical water calculations at the RHF/STO-3G level using different mixing schemes and self-energy strengths $\eta$. 
    Left: calculations initialized from the converged canonical ($\boldsymbol{\Sigma}=0$) density matrix. 
    Right: calculations initialized from the SAD guess. 
    The panels show the convergence of the grand potential difference $|\Delta \Omega|$, the density-matrix change $\|\Delta \mathbf{P}\|$, and the number of electrons $N_e$ as a function of SCF iteration.
    }
    \label{fig:water_conv_rhf_sto3g}
\end{figure}


\begin{table}
\caption{Final grand-canonical SCF results for the water molecule using the \textbf{MINAO} basis at \textbf{RHF} level. Reported are the grand potential $\Omega$, the number of electrons $N_e$, and the Frobenius norm of the FPS residual $\|\mathbf{R}\|_F$. Residuals are reported only for converged calculations; $\dagger$ marks a run that did not meet the convergence criterion. The reference chemical potential is $\mu_{\mathrm{ref}} = -0.08565~\mathrm{Ha}$.}
\label{tab:water_scf_results_log}
\centering
\begin{tabular}{l|cccc|cccc}
\hline
& \multicolumn{4}{c|}{SAD guess} & \multicolumn{4}{c}{$\boldsymbol{\Sigma}=0$ start} \\
SCF scheme & $\Omega$ (Ha) & $N_e$ & iter & $\|\mathbf{R}\|_F$ & $\Omega$ (Ha) & $N_e$ & iter & $\|\mathbf{R}\|_F$ \\
\hline
\multicolumn{9}{c}{$\eta = 1.0\times10^{-6}$ Ha} \\
\hline
EDIIS(E) & -75.05214 & 10.00000 & 80 & $2.61\times10^{-6}$ & -75.05214 & 10.00000 & 3 & $1.71\times10^{-6}$ \\
CDIIS & -75.05214 & 10.00000 & 9 & $1.68\times10^{-6}$ & -75.05214 & 10.00000 & 3 & $1.72\times10^{-6}$ \\
Anderson & -75.05214 & 10.00000 & 19 & $1.74\times10^{-6}$ & -75.05214 & 10.00000 & 40 & $1.72\times10^{-6}$ \\
EDIIS($\Omega$) & -75.05214 & 10.00000 & 80 & $2.61\times10^{-6}$ & -75.05214 & 10.00000 & 3 & $1.71\times10^{-6}$ \\
EDIIS($\Omega$)+A & -75.05214 & 10.00000 & 26 & $1.73\times10^{-6}$ & -75.05214 & 10.00000 & 3 & $1.71\times10^{-6}$ \\
\hline
\multicolumn{9}{c}{$\eta = 5\times10^{-3}$ Ha} \\
\hline
EDIIS(E) & -75.02239 & 10.01520 & 80 & $8.56\times10^{-3}$ & -75.02239 & 10.01520 & 47 & $8.56\times10^{-3}$ \\
CDIIS & -75.00796$^\dagger$ & 10.01510 & 201 & -- & -75.02246 & 10.01564 & 3 & $8.51\times10^{-3}$ \\
Anderson & -75.02239 & 10.01520 & 20 & $8.56\times10^{-3}$ & -75.02239 & 10.01520 & 78 & $8.56\times10^{-3}$ \\
EDIIS($\Omega$) & -75.02239 & 10.01520 & 80 & $8.56\times10^{-3}$ & -75.02239 & 10.01520 & 47 & $8.56\times10^{-3}$ \\
EDIIS($\Omega$)+A & -75.02239 & 10.01520 & 26 & $8.56\times10^{-3}$ & -75.02239 & 10.01520 & 38 & $8.56\times10^{-3}$ \\
\hline
\multicolumn{9}{c}{$\eta = 1\times10^{-1}$ Ha} \\
\hline
EDIIS(E) & -74.47458 & 10.17987 & 78 & $1.39\times10^{-1}$ & -74.47458 & 10.17988 & 70 & $1.39\times10^{-1}$ \\
CDIIS & -71.11016$^\dagger$ & 13.34391 & 201 & -- & -74.50850 & 10.25822 & 3 & $1.40\times10^{-1}$ \\
Anderson & -74.47458 & 10.17988 & 76 & $1.39\times10^{-1}$ & -74.47458 & 10.17988 & 117 & $1.39\times10^{-1}$ \\
EDIIS($\Omega$) & -74.47458 & 10.17987 & 78 & $1.39\times10^{-1}$ & -74.47458 & 10.17988 & 70 & $1.39\times10^{-1}$ \\
EDIIS($\Omega$)+A & -74.47458 & 10.17988 & 28 & $1.39\times10^{-1}$ & -74.47458 & 10.17988 & 47 & $1.39\times10^{-1}$ \\
\hline
\end{tabular}
\end{table}

\begin{figure}[H]
    \centering

    \begin{minipage}{0.49\textwidth}
        \centering
        \includegraphics[width=\linewidth]
        {Figures/SI Figures/water_convergence__rhf__basis-minao__start-canonical_same_theory__density-log.png}
    \end{minipage}
    \hfill
    \begin{minipage}{0.49\textwidth}
        \centering
        \includegraphics[width=\linewidth]
        {Figures/SI Figures/water_convergence__rhf__basis-minao__start-sad__density-log.png}
    \end{minipage}

    \caption{
    SCF convergence behavior for the grand-canonical water calculations at the RHF/MINAO level using different mixing schemes and self-energy strengths $\eta$. 
    Left: calculations initialized from the converged canonical ($\boldsymbol{\Sigma}=0$) density matrix. 
    Right: calculations initialized from the SAD guess. 
    The panels show the convergence of the grand potential difference $|\Delta \Omega|$, the density-matrix change $\|\Delta \mathbf{P}\|$, and the number of electrons $N_e$ as a function of SCF iteration.
    }
    \label{fig:water_conv_rhf_minao}
\end{figure}

\newpage
\subsection{S2.2 Density Functional Theory}
\addcontentsline{toc}{subsection}{S2.2 Density Functional Theory}
\begin{table}
\caption{Final grand-canonical SCF results for the water molecule using the \textbf{STO-3G} basis at \textbf{RKS/PBE} level. Reported are the grand potential $\Omega$, the number of electrons $N_e$, and the Frobenius norm of the FPS residual $\|\mathbf{R}\|_F$. Residuals are reported only for converged calculations; $\dagger$ marks a run that did not meet the convergence criterion. The reference chemical potential is $\mu_{\mathrm{ref}} = 0.10693~\mathrm{Ha}$.}
\label{tab:water_scf_results_log}
\centering
\begin{tabular}{l|cccc|cccc}
\hline
& \multicolumn{4}{c|}{SAD guess} & \multicolumn{4}{c}{$\boldsymbol{\Sigma}=0$ start} \\
SCF scheme & $\Omega$ (Ha) & $N_e$ & iter & $\|\mathbf{R}\|_F$ & $\Omega$ (Ha) & $N_e$ & iter & $\|\mathbf{R}\|_F$ \\
\hline
\multicolumn{9}{c}{$\eta = 1.0\times10^{-6}$ Ha} \\
\hline
EDIIS(E) & -76.29495 & 10.00001 & 64 & $3.34\times10^{-5}$ & -76.29495 & 10.00001 & 4 & $1.49\times10^{-6}$ \\
CDIIS & -76.29495 & 10.00001 & 8 & $1.66\times10^{-6}$ & -76.29495 & 10.00001 & 3 & $1.73\times10^{-6}$ \\
Anderson & -76.29495 & 10.00001 & 92 & $2.86\times10^{-6}$ & -76.29495 & 10.00001 & 37 & $1.54\times10^{-6}$ \\
EDIIS($\Omega$) & -76.29495 & 10.00001 & 64 & $3.34\times10^{-5}$ & -76.29495 & 10.00001 & 4 & $1.49\times10^{-6}$ \\
EDIIS($\Omega$)+A & -76.29495 & 10.00001 & 29 & $1.54\times10^{-6}$ & -76.29495 & 10.00001 & 4 & $1.49\times10^{-6}$ \\
\hline
\multicolumn{9}{c}{$\eta = 5\times10^{-3}$ Ha} \\
\hline
EDIIS(E) & -76.26793 & 10.02174 & 80 & $7.59\times10^{-3}$ & -76.26793 & 10.02174 & 35 & $7.59\times10^{-3}$ \\
CDIIS & -70.87906$^\dagger$ & 13.99479 & 201 & -- & -76.26753 & 10.02571 & 4 & $8.22\times10^{-3}$ \\
Anderson & -76.26793 & 10.02174 & 95 & $7.59\times10^{-3}$ & -76.26793 & 10.02174 & 79 & $7.59\times10^{-3}$ \\
EDIIS($\Omega$) & -76.26793 & 10.02174 & 80 & $7.59\times10^{-3}$ & -76.26793 & 10.02174 & 35 & $7.59\times10^{-3}$ \\
EDIIS($\Omega$)+A & -76.26793 & 10.02174 & 35 & $7.59\times10^{-3}$ & -76.26793 & 10.02174 & 26 & $7.59\times10^{-3}$ \\
\hline
\multicolumn{9}{c}{$\eta = 1\times10^{-1}$ Ha} \\
\hline
EDIIS(E) & -75.83945 & 10.13849 & 201 & $1.08\times10^{-1}$ & -75.83945 & 10.13846 & 93 & $1.08\times10^{-1}$ \\
CDIIS & -52.56482$^\dagger$ & 2.20011 & 201 & -- & -75.79293 & 10.27154 & 4 & $1.34\times10^{-1}$ \\
Anderson & -75.83945 & 10.13847 & 107 & $1.08\times10^{-1}$ & -75.83945 & 10.13847 & 105 & $1.08\times10^{-1}$ \\
EDIIS($\Omega$) & -75.83945 & 10.13849 & 201 & $1.08\times10^{-1}$ & -75.83945 & 10.13846 & 93 & $1.08\times10^{-1}$ \\
EDIIS($\Omega$)+A & -75.83945 & 10.13847 & 38 & $1.08\times10^{-1}$ & -75.83945 & 10.13847 & 34 & $1.08\times10^{-1}$ \\
\hline
\end{tabular}
\end{table}

\begin{figure}[H]
    \centering

    \begin{minipage}{0.49\textwidth}
        \centering
        \includegraphics[width=\linewidth]
        {Figures/SI Figures/water_convergence__rks_pbe__basis-sto-3g__start-canonical_same_theory__density-log.png}
    \end{minipage}
    \hfill
    \begin{minipage}{0.49\textwidth}
        \centering
        \includegraphics[width=\linewidth]
        {Figures/SI Figures/water_convergence__rks_pbe__basis-sto-3g__start-sad__density-log.png}
    \end{minipage}

    \caption{
    SCF convergence behavior for the grand-canonical water calculations at the PBE/STO-3G level using different mixing schemes and self-energy strengths $\eta$. 
    Left: calculations initialized from the converged canonical ($\boldsymbol{\Sigma}=0$) density matrix. 
    Right: calculations initialized from the SAD guess. 
    The panels show the convergence of the grand potential difference $|\Delta \Omega|$, the density-matrix change $\|\Delta \mathbf{P}\|$, and the number of electrons $N_e$ as a function of SCF iteration.
    }
\label{fig:water_conv_pbe_sto3g}
\end{figure}

\begin{table}
\caption{Final grand-canonical SCF results for the water molecule using the \textbf{MINAO} basis at \textbf{RKS/PBE} level. Reported are the grand potential $\Omega$, the number of electrons $N_e$, and the Frobenius norm of the FPS residual $\|\mathbf{R}\|_F$. Residuals are reported only for converged calculations; $\dagger$ marks a run that did not meet the convergence criterion. The reference chemical potential is $\mu_{\mathrm{ref}} = -0.08565~\mathrm{Ha}$.}
\label{tab:water_scf_results_log}
\centering
\begin{tabular}{l|cccc|cccc}
\hline
& \multicolumn{4}{c|}{SAD guess} & \multicolumn{4}{c}{$\boldsymbol{\Sigma}=0$ start} \\
SCF scheme & $\Omega$ (Ha) & $N_e$ & iter & $\|\mathbf{R}\|_F$ & $\Omega$ (Ha) & $N_e$ & iter & $\|\mathbf{R}\|_F$ \\
\hline
\multicolumn{9}{c}{$\eta = 1.0\times10^{-6}$ Ha} \\
\hline
EDIIS(E) & -75.36468 & 10.00001 & 72 & $1.15\times10^{-5}$ & -75.36468 & 10.00001 & 4 & $1.77\times10^{-6}$ \\
CDIIS & -75.36468 & 10.00001 & 8 & $1.93\times10^{-6}$ & -75.36468 & 10.00001 & 3 & $1.81\times10^{-6}$ \\
Anderson & -75.36468 & 10.00001 & 94 & $2.61\times10^{-6}$ & -75.36468 & 10.00001 & 36 & $1.82\times10^{-6}$ \\
EDIIS($\Omega$) & -75.36468 & 10.00001 & 72 & $1.15\times10^{-5}$ & -75.36468 & 10.00001 & 4 & $1.77\times10^{-6}$ \\
EDIIS($\Omega$)+A & -75.36468 & 10.00001 & 31 & $1.80\times10^{-6}$ & -75.36468 & 10.00001 & 4 & $1.77\times10^{-6}$ \\
\hline
\multicolumn{9}{c}{$\eta = 5\times10^{-3}$ Ha} \\
\hline
EDIIS(E) & -75.33519 & 10.02604 & 63 & $8.96\times10^{-3}$ & -75.33519 & 10.02604 & 35 & $8.94\times10^{-3}$ \\
CDIIS & -51.93512$^\dagger$ & 2.00555 & 201 & -- & -75.33476 & 10.03086 & 3 & $8.85\times10^{-3}$ \\
Anderson & -75.33519 & 10.02604 & 97 & $8.94\times10^{-3}$ & -75.33519 & 10.02604 & 73 & $8.94\times10^{-3}$ \\
EDIIS($\Omega$) & -75.33519 & 10.02604 & 63 & $8.96\times10^{-3}$ & -75.33519 & 10.02604 & 35 & $8.94\times10^{-3}$ \\
EDIIS($\Omega$)+A & -75.33519 & 10.02604 & 35 & $8.94\times10^{-3}$ & -75.33519 & 10.02604 & 26 & $8.94\times10^{-3}$ \\
\hline
\multicolumn{9}{c}{$\eta = 1\times10^{-1}$ Ha} \\
\hline
EDIIS(E) & -74.88930 & 10.17337 & 125 & $1.26\times10^{-1}$ & -74.88927 & 10.17283 & 105 & $1.27\times10^{-1}$ \\
CDIIS & -52.15567$^\dagger$ & 2.06640 & 201 & -- & -74.84537 & 10.28004 & 4 & $1.50\times10^{-1}$ \\
Anderson & -74.88927 & 10.17283 & 106 & $1.27\times10^{-1}$ & -74.88927 & 10.17283 & 104 & $1.27\times10^{-1}$ \\
EDIIS($\Omega$) & -74.88928 & 10.17291 & 184 & $1.27\times10^{-1}$ & -74.88927 & 10.17283 & 105 & $1.27\times10^{-1}$ \\
EDIIS($\Omega$)+A & -74.88927 & 10.17283 & 37 & $1.27\times10^{-1}$ & -74.88927 & 10.17283 & 35 & $1.27\times10^{-1}$ \\
\hline
\end{tabular}
\end{table}

\begin{figure}[H]
    \centering

    \begin{minipage}{0.49\textwidth}
        \centering
        \includegraphics[width=\linewidth]
        {Figures/SI Figures/water_convergence__rks_pbe__basis-minao__start-canonical_same_theory__density-log.png}
    \end{minipage}
    \hfill
    \begin{minipage}{0.49\textwidth}
        \centering
        \includegraphics[width=\linewidth]
        {Figures/SI Figures/water_convergence__rks_pbe__basis-minao__start-sad__density-log.png}
    \end{minipage}

    \caption{
    SCF convergence behavior for the grand-canonical water calculations at the PBE/MINAO level using different mixing schemes and self-energy strengths $\eta$. 
    Left: calculations initialized from the converged canonical ($\boldsymbol{\Sigma}=0$) density matrix. 
    Right: calculations initialized from the SAD guess. 
    The panels show the convergence of the grand potential difference $|\Delta \Omega|$, the density-matrix change $\|\Delta \mathbf{P}\|$, and the number of electrons $N_e$ as a function of SCF iteration.
    }
  \label{fig:water_conv_pbe_minao}
\end{figure}

\newpage
\section{S3 Further information and results for the 1D quasi-periodic system}
\addcontentsline{toc}{section}{S3 Further information and results for the 1D quasi-periodic system}
In this section, we present additional results and information on the hydrogen ring system, including coordinates, optimization protocols, and results investigating its metallic behavior.
\subsection{S3.1 Coordinates of the hydrogen ring}
\addcontentsline{toc}{subsection}{S3.1 Coordinates of the hydrogen ring}
The coordinates of the full ring \ce{H40}, the two fragments \ce{H6} and \ce{H38} are shown below (in Å).\\

\begin{figure}[H]
\centering

\begin{tcolorbox}[
  colback=gray!10,
  colframe=gray!40,
  boxrule=0.3pt,
  arc=2pt,
  width=\textwidth
]
\begin{lstlisting}[basicstyle=\ttfamily\tiny]
40

H -4.701296 -0.370000 0.000000
H -4.585534 -1.100889 0.000000
H -4.356862 -1.804671 0.000000
H -4.020909 -2.464016 0.000000
H -3.585948 -3.062689 0.000000
H -3.062689 -3.585948 0.000000
H -2.464016 -4.020909 0.000000
H -1.804671 -4.356862 0.000000
H -1.100889 -4.585534 0.000000
H -0.370000 -4.701296 0.000000
H 0.370000 -4.701296 0.000000
H 1.100889 -4.585534 0.000000
H 1.804671 -4.356862 0.000000
H 2.464016 -4.020909 0.000000
H 3.062689 -3.585948 0.000000
H 3.585948 -3.062689 0.000000
H 4.020909 -2.464016 0.000000
H 4.356862 -1.804671 0.000000
H 4.585534 -1.100889 0.000000
H 4.701296 -0.370000 0.000000
H 4.701296 0.370000 0.000000
H 4.585534 1.100889 0.000000
H 4.356862 1.804671 0.000000
H 4.020909 2.464016 0.000000
H 3.585948 3.062689 0.000000
H 3.062689 3.585948 0.000000
H 2.464016 4.020909 0.000000
H 1.804671 4.356862 0.000000
H 1.100889 4.585534 0.000000
H 0.370000 4.701296 0.000000
H -0.370000 4.701296 0.000000
H -1.100889 4.585534 0.000000
H -1.804671 4.356862 0.000000
H -2.464016 4.020909 0.000000
H -3.062689 3.585948 0.000000
H -3.585948 3.062689 0.000000
H -4.020909 2.464016 0.000000
H -4.356862 1.804671 0.000000
H -4.585534 1.100889 0.000000
H -4.701296 0.370000 0.000000
\end{lstlisting}
\end{tcolorbox}

\caption{Cartesian coordinates (in Å) of the H$_{40}$ ring used in this work.}
\label{fig:h40_xyz}

\end{figure}

\begin{figure}[H]
\centering

\begin{tcolorbox}[
  colback=gray!10,
  colframe=gray!40,
  boxrule=0.3pt,
  arc=2pt,
  width=\textwidth
]
\begin{lstlisting}[basicstyle=\ttfamily\tiny]
38

H -4.585534 -1.100889 0.000000
H -4.356862 -1.804671 0.000000
H -4.020909 -2.464016 0.000000
H -3.585948 -3.062689 0.000000
H -3.062689 -3.585948 0.000000
H -2.464016 -4.020909 0.000000
H -1.804671 -4.356862 0.000000
H -1.100889 -4.585534 0.000000
H -0.370000 -4.701296 0.000000
H 0.370000 -4.701296 0.000000
H 1.100889 -4.585534 0.000000
H 1.804671 -4.356862 0.000000
H 2.464016 -4.020909 0.000000
H 3.062689 -3.585948 0.000000
H 3.585948 -3.062689 0.000000
H 4.020909 -2.464016 0.000000
H 4.356862 -1.804671 0.000000
H 4.585534 -1.100889 0.000000
H 4.701296 -0.370000 0.000000
H 4.701296 0.370000 0.000000
H 4.585534 1.100889 0.000000
H 4.356862 1.804671 0.000000
H 4.020909 2.464016 0.000000
H 3.585948 3.062689 0.000000
H 3.062689 3.585948 0.000000
H 2.464016 4.020909 0.000000
H 1.804671 4.356862 0.000000
H 1.100889 4.585534 0.000000
H 0.370000 4.701296 0.000000
H -0.370000 4.701296 0.000000
H -1.100889 4.585534 0.000000
H -1.804671 4.356862 0.000000
H -2.464016 4.020909 0.000000
H -3.062689 3.585948 0.000000
H -3.585948 3.062689 0.000000
H -4.020909 2.464016 0.000000
H -4.356862 1.804671 0.000000
H -4.585534 1.100889 0.000000
\end{lstlisting}
\end{tcolorbox}

\caption{Cartesian coordinates (in Å) of the H$_{38}$ ring used in this work.}
\label{fig:h38_xyz}

\end{figure}

\begin{figure}[H]
\centering

\begin{tcolorbox}[
  colback=gray!10,
  colframe=gray!40,
  boxrule=0.3pt,
  arc=2pt,
  width=\textwidth
]
\begin{lstlisting}[basicstyle=\ttfamily\tiny]
6

H 4.356862 -1.804671 0.000000
H 4.585534 -1.100889 0.000000
H 4.701296 -0.370000 0.000000
H 4.701296 0.370000 0.000000
H 4.585534 1.100889 0.000000
H 4.356862 1.804671 0.000000

\end{lstlisting}
\end{tcolorbox}

\caption{Cartesian coordinates (in Å) of the H$_{6}$ ring used in this work.}
\label{fig:h6_xyz}

\end{figure}

\subsection{S3.2 Optimization protocol and Landscape of the objective}
\addcontentsline{toc}{subsection}{S3.2 Optimization protocol}

The loss function used for the optimization of the coupling parameters computes the point-wise mean squared
error (MSE) between the grand-canonical fragment interaction curve, $\Delta\Omega_{\mathrm{frag}}(r)$, and the
reference interaction curve of the full hydrogen ring, $\Delta\Omega_{\mathrm{ref}}(r)$. Both curves are shifted
relative to the respective system with the point charge fully removed ($q=0$),
\begin{equation}
\Delta\Omega(r) = \Omega(r) - \Omega(q=0).
\end{equation}

The objective function is then defined as
\begin{equation}
\mathcal{L}
=
\frac{1}{N}
\sum_{i=1}^{N}
\left[
\Delta\Omega_{\mathrm{frag}}(r_i)
-
\Delta\Omega_{\mathrm{ref}}(r_i)
\right]^2
\end{equation}
where $N$ is the number of scan points.

The interaction curve was evaluated on the same probe distance grid as used for the full-ring reference calculation. The total number of points and the corresponding distances are listed in Table~S\ref{tab:scan_grid}. All fragment single-point calculations were performed at each probe position for a given parameter set $(\epsilon, \eta)$.

\begin{table}
\caption{Probe--ring separation distances $r$ (in \AA) used for the point-charge scan.}
\label{tab:scan_grid}
\centering
\begin{tabular}{cccccccc}
\hline
$r$ (\AA) && &  &  & &  & \\
\hline
0.250 & 0.375 & 0.500 & 0.625 & 0.750 & 0.875 & 1.000 & 1.125 \\
1.250 & 1.375 & 1.500 & 1.625 & 1.750 & 1.875 & 2.000 & 2.125 \\
2.250 & 2.375 & 2.500 & 2.625 & 2.750 & 2.875 & 3.000 & 3.125 \\
3.250 & 3.375 & 3.500 & 3.750 & 4.000 & 4.250 & 4.500 & 4.750 \\
5.000 & 5.250 & 5.500 & 5.750 & 6.000 & 6.250 & 6.500 & 6.750 \\
7.000 & 7.250 & 8.000 & 8.500 & 9.000 & 9.500 & 10.000 & 10.500 \\
11.000 & 11.500 & 12.000 & 12.500 & 13.000 & 13.500 & 14.000 & 14.500 \\
15.000 & 15.500 & 16.000 & 16.500 & 17.000 & 17.500 & 18.000 & 18.500 \\
19.000 & 19.500 & 50.000 & 100.000 &  &  &  &  \\
\hline
\end{tabular}
\end{table}
To determine the chemical potential under the neutrality constraint, the fragment was required to remain neutral upon removal of the point charge, i.e., $N_e^{(0)} = 6$. For each trial pair $(\epsilon, \eta)$, the chemical potential $\mu$ was adjusted by solving
\begin{equation}
N_e(\mu) - N_e^{(0)} = 0
\end{equation}
using Brent's root-finding method\cite{brent2013algorithms} as implemented in \texttt{scipy}\cite{2020SciPy-NMeth}. \\
Figure~\ref{fig:contour_scan} shows the loss function and the corresponding neutrality-constrained $\mu$ over $\varepsilon \in [-0.30, 0.15]$~Ha and $\eta \in [0.02, 0.35]$~Ha.
\begin{figure}[H]
    \centering

        \includegraphics[width=\linewidth]{Figures/SI Figures/heatmap_objective_and_mu_curve_minao_rks_pbe0_H6.png}
  \caption{Grid scan of the loss function (left) and the resulting chemical
potential $\mu$ under the neutrality constraint (right) over the coupling
parameters $\varepsilon$ and $\eta$.}
\label{fig:contour_scan}
\end{figure}

\begin{figure}[H]
    \centering

        \includegraphics[width=\linewidth]{Figures/SI Figures/Hring_second_minimum.png}
  \caption{\textbf{(a)} Potential curve of the point charge approaching the hydrogen system for the uncoupled fragment (blue), the reference system, and the coupled systems with the optimized parameters shown in the upper panel, where $\Delta \Omega=\Omega(r)-\Omega_0$, where $\Omega_0$ is the grand potential of the fragment in the absence of a positive point charge. The lower panel shows the residual relative to the \ce{H40} reference system in kcal/mol. The obtained chemical potential fulfilling the neutrality constraint is $-0.2767$ Ha.
\textbf{(b)}
Two-dimensional $xy$-plane ($z=0$) electron-density slices at the adsorption minimum ($r=1.22~\text{\AA}$) for an external point charge $q=+1$ placed along the ring axis (white cross). See Fig.~3 for more details.
\label{fig:hring}
}
\label{fig:second_minimum}
\end{figure}

\newpage

\subsection{S3.3 Further results for all optimized fragment sizes}
\addcontentsline{toc}{subsection}{S3.3 Further results for all optimized fragment sizes}

Figure~\ref{fig:overlay_convergence} shows the resulting adsorption potential for each fragment size.
\begin{figure}[H]
\includegraphics[width=0.5\linewidth]{Figures/SI Figures/fragment_convergence_log.png}
\caption{
Convergence of the potential energy curve with fragment size. The upper panel shows the potential energy curves obtained for the optimized GC-SCF fragments ranging from \ce{H6} to \ce{H20}, together with the \ce{H40} reference and the corresponding uncoupled fragments. The lower panel shows the residuals relative to the \ce{H40} reference.}\label{fig:overlay_convergence}
\end{figure}

Figure~\ref{fig:density_convergence_si} shows density-difference plots at the adsorption minimum for the optimized fragments discussed in the convergence study. Figure~\ref{fig:overlay_convergence} shows the resulting adsorption potential for each fragment size.

\begin{figure}[H]
\centering

\begin{minipage}{0.48\textwidth}
    \centering
    \includegraphics[width=\linewidth]{Figures/SI Figures/density_comparison__H6__eps0m0.0655__gamma00.2048__paper.png}
    \end{minipage}
\hfill
\begin{minipage}{0.48\textwidth}
    \centering
    \includegraphics[width=\linewidth]{Figures/SI Figures/density_comparison__H8_a__eps0m0.0594__gamma00.1574__paper.png}
\end{minipage}
\\[0.5em]
\begin{minipage}{0.48\textwidth}
    \centering
    \includegraphics[width=\linewidth]{Figures/SI Figures/density_comparison__H10__eps0m0.0608__gamma00.1345__paper.png}
    \end{minipage}
\hfill
\begin{minipage}{0.48\textwidth}
    \centering
    \includegraphics[width=\linewidth]{Figures/SI Figures/density_comparison__H12__eps0m0.0625__gamma00.1343__paper.png}
\end{minipage}    
\\[0.5em]
\begin{minipage}{0.48\textwidth}
    \centering
    \includegraphics[width=\linewidth]{Figures/SI Figures/density_comparison__H14__eps0m0.0413__gamma00.0862__paper.png}
\end{minipage}
\hfill
\begin{minipage}{0.48\textwidth}
    \centering
    \includegraphics[width=\linewidth]{Figures/SI Figures/density_comparison__H16__eps0m0.0457__gamma00.0896__paper.png}  
\end{minipage}
\begin{minipage}{0.48\textwidth}
    \centering
    \includegraphics[width=\linewidth]{Figures/SI Figures/density_comparison__H18__eps0m0.0246__gamma00.0599__paper.png}
\end{minipage}
\hfill
\begin{minipage}{0.48\textwidth}
    \centering
    \includegraphics[width=\linewidth]{Figures/SI Figures/density_comparison__H20__eps0m0.0262__gamma00.0597__paper.png}  
\end{minipage}
\caption{
Density-difference plots at the adsorption minimum for the optimized fragments of differing size as discussed in the convergence study.
}
\label{fig:density_convergence_si}
\end{figure}

\subsection{S3.4 Electronic-structure analysis and effective temperature}
\addcontentsline{toc}{subsection}{S3.4 Electronic-structure analysis and effective temperature}
To quantify the occupation broadening induced by the grand-canonical treatment, we extract an effective electronic temperature from the fractional orbital occupations of the fragment. 

In the grand-canonical formalism, the fragment orbitals acquire fractional occupations $n_i$ in the vicinity of the chemical potential $\mu$. 
To characterize the width of this transition region, we fit the occupations to a Fermi--Dirac distribution. Although the applied broadening leads to a Lorentzian distribution, for better comparability we fit a Fermi--Dirac distribution of the form

\begin{equation}
n_i \approx \frac{2}{1 + \exp\left[\beta(\varepsilon_i - \mu)\right]},
\end{equation}

where $\varepsilon_i$ are the orbital energies, $\mu$ is the chemical potential used in the calculation, and $\beta = 1/(k_\mathrm{B}T_\mathrm{eff})$ defines an effective electronic temperature $T_\mathrm{eff}$.

Rearranging the above expression yields a linear relation,

\begin{equation}
\ln\!\left(\frac{2}{n_i} - 1\right) = \beta (\varepsilon_i - \mu),
\end{equation}

which allows $\beta$ to be obtained from a least-squares fit of 
$y_i = \ln\!\left(\frac{2}{n_i} - 1\right)$ 
against 
$x_i = \varepsilon_i - \mu$.
The fit is restricted to orbitals with fractional occupations (i.e., excluding occupations close to 0 or 2) and to a small energy window around $\mu$ to avoid contributions from fully occupied or empty states. The effective temperature is then obtained as

\begin{equation}
T_\mathrm{eff} = \frac{1}{k_\mathrm{B} \beta}.
\end{equation}

\begin{table}[H]
\caption{Canonical HOMO--LUMO gaps, chemical-potential-centered frontier orbital spacings, and effective electronic temperatures. The quantity $\Delta_\mu$ is defined as the energy spacing between the closest grand-canonical orbital below and above the chemical potential. The effective temperature is a diagnostic measure of occupation broadening and does not correspond to a physical temperature.}
\label{tab:gaps_teff}
\centering
\begin{tabular}{lccccccc}
\hline
System & $\Delta E_\mathrm{HL}$ & $\Delta E_\mathrm{HL}(q@r_\mathrm{min})$ & $\Delta_\mu$ & $\Delta_\mu(q@r_\mathrm{min})$ & $T_\mathrm{eff}$ & $T_\mathrm{eff}(q@r_\mathrm{min})$ & $k_\mathrm{B}T_\mathrm{eff}/\Delta_\mu$ \\
 & (Ha) & (Ha) & (Ha) & (Ha) & (K) & (K) & \\
\hline
H$_{6}$ & 0.385346 & 0.390487 & 0.301971 & 0.247776 & 31\,907 & 31\,862 & 0.335 \\
H$_{8}$ & 0.319719 & 0.316219 & 0.268652 & 0.238599 & 28\,699 & 33\,808 & 0.338 \\
H$_{10}$ & 0.274393 & 0.276353 & 0.203982 & 0.220964 & 24\,656 & 26\,720 & 0.383 \\
H$_{12}$ & 0.241048 & 0.237884 & 0.173547 & 0.188510 & 22\,901 & 27\,414 & 0.418 \\
H$_{14}$ & 0.215419 & 0.216502 & 0.177720 & 0.186781 & 19\,215 & 27\,472 & 0.342 \\
H$_{16}$ & 0.195074 & 0.192401 & 0.158674 & 0.163788 & 18\,749 & 27\,150 & 0.374 \\
H$_{18}$ & 0.178515 & 0.179270 & 0.159689 & 0.163777 & 19\,762 & 24\,461 & 0.392 \\
H$_{20}$ & 0.164770 & 0.162491 & 0.146521 & 0.147828 & 18\,652 & 23\,404 & 0.403 \\
H$_{38}$ & 0.114780 & 0.114600 & -- & -- & -- & -- & -- \\
H$_{40}$ & 0.037560 & 0.037649 & -- & -- & -- & -- & -- \\
\hline
\end{tabular}
\end{table}

\begin{figure}[H]
    \centering

        \includegraphics[width=0.5\linewidth]{Figures/SI Figures/mu_alignment__paper.png}
    \caption{Chemical potentials obtained from the neutrality constraint for the optimized hydrogen-ring fragments. Shown are the HOMO and LUMO energies of the coupled and uncoupled fragments together with the chemical potential $\mu$ obtained from the neutrality-constrained grand-canonical optimizations.}
\label{fig:mu_alignemt}
\end{figure}

\newpage

\section{S4 Further information and results for the Li cluster}
\addcontentsline{toc}{section}{S4 Further information and results for the Li cluster}

\subsection{S4.1 Coordinates and coupling schemes of the Li cluster}
\addcontentsline{toc}{subsection}{S4.1 Coordinates of Li cluster}
The coordinates of the \ce{Li40} bcc cluster from the 110 facet are shown below (in Å). Moreover, the indices of the atoms in the differing coupling schemes are shown in Table~\ref{tab:coupling}.\\

\begin{figure}[H]
\centering

\begin{tcolorbox}[
  colback=gray!10,
  colframe=gray!40,
  boxrule=0.3pt,
  arc=2pt,
  width=\textwidth
]
\begin{lstlisting}[basicstyle=\ttfamily\tiny]
40

Li         7.44583        5.26500       12.00000
Li         7.44583        3.51000       14.48194
Li         4.96389        5.26500       14.48194
Li         7.44583        5.26500       16.96389
Li         7.44583        8.77500       12.00000
Li         7.44583        7.02000       14.48194
Li         4.96389        8.77500       14.48194
Li         7.44583        8.77500       16.96389
Li         7.44583       10.53000       14.48194
Li         9.92778        1.75500       14.48194
Li         9.92778        3.51000       12.00000
Li        12.40972        5.26500       12.00000
Li        12.40972        3.51000       14.48194
Li         9.92778        5.26500       14.48194
Li         9.92778        3.51000       16.96389
Li        12.40972        5.26500       16.96389
Li         9.92778        7.02000       12.00000
Li        12.40972        8.77500       12.00000
Li        12.40972        7.02000       14.48194
Li         9.92778        8.77500       14.48194
Li         9.92778        7.02000       16.96389
Li        12.40972        8.77500       16.96389
Li         9.92778       10.53000       12.00000
Li        12.40972       10.53000       14.48194
Li         9.92778       12.28500       14.48194
Li         9.92778       10.53000       16.96389
Li        14.89167        5.26500       14.48194
Li        14.89167        8.77500       14.48194
Li         9.92779        0.00000       12.00000
Li         7.44584        1.75500       12.00000
Li         4.96389        3.51000       12.00000
Li        12.40973        1.75500       12.00000
Li         4.96388        7.02000       12.00000
Li        14.89167        3.51000       12.00000
Li         4.96387       10.53000       12.00000
Li        14.89167        7.02000       12.00000
Li         7.44582       12.28500       12.00000
Li        14.89166       10.53000       12.00000
Li        12.40971       12.28500       12.00000
Li         9.92776       14.04000       12.00000
\end{lstlisting}
\end{tcolorbox}

\caption{Cartesian coordinates (in Å) of the Li$_{40}(110)$ cluster used in this work.}
\label{fig:li_cluster_coords}

\end{figure}
\begin{table}
\caption{Atom indices according to \ref{fig:li_cluster_coords} for the different coupling topologies.}\label{tab:coupling}
\centering
\begin{tabular}{lcp{9cm}}
\hline
Scheme & $N_\mathrm{coupled}$ & Atom indices \\
\hline
Center         & 4  & 6, 14, 19, 20 \\
Surface center & 7  & 1, 5, 11, 12, 17, 18, 23 \\
Edges          & 29 & 2, 3, 4, 7, 8, 9, 10, 13, 15, 16, 21, 22, 24, 25, 26, 27, 28, 29, 30, 31, 32, 33, 34, 35, 36, 37, 38, 39, 40 \\
\hline
\end{tabular}
\end{table}

\subsection{S4.2 Further results for the differing coupling schemes}
\addcontentsline{toc}{subsection}{S4.2 Further results for the differing coupling schemes}

%

\begin{figure}[H]
    \centering

        \includegraphics[width=1.0\linewidth]{Figures/SI Figures/broadening_SI.png}
    \caption{MO energies $e_i$ (upper panel) and
grand-canonical occupations $f_i$
(lower panel) as a function of the broadening $\eta$ for the
\ce{Li40} cluster coupled at the seven surface center atoms (left) and the 29 edge atoms (right).
Blue curves show the three highest occupied orbitals
and red curves the
three lowest unoccupied orbitals.
Dashed horizontal lines mark the canonical HOMO and LUMO
energies where the self-consistent chemical potential $\mu(\eta)$ is
shown in black. The vertical dashed line indicates the canonical
HOMO--LUMO gap (24.5~mHa).}
\end{figure}

\begin{figure}[H]
    \centering

        \includegraphics[width=0.5\linewidth]{Figures/SI Figures/fig_density_scaling_pbe0_g5e-03.png}
    \caption{ Integrated absolute density perturbation as a function of the number of GC-coupled atoms at $\eta = 5,\mathrm{mHa}$ (PBE0/STO-3G). The three data points correspond to the center (4 atoms), surface-center (7 atoms), and edges (29 atoms) coupling schemes. Dashed lines show linear fits through the origin.}
\label{fig:scaling}
\end{figure}

\newpage

\bibliography{refs}